\newif\ifprocs
\procstrue
\procsfalse 

\newif\ifarxiv
\arxivtrue

\newif\ifcomments
\commentstrue

\ifprocs 
\documentclass[twoside,leqno,twocolumn]{article}
\usepackage[letterpaper]{geometry}
\usepackage{ltexpprt}

\else
\documentclass[11pt,USenglish]{article}
\usepackage[margin=1.0in]{geometry}
\usepackage[T1]{fontenc}
\usepackage[utf8]{inputenc}
\usepackage[english]{babel}
\usepackage{authblk}
\usepackage{chngcntr}
\usepackage[T1]{fontenc}
\counterwithin{equation}{section}
\fi
\usepackage[normalem]{ulem}
\usepackage{xfrac}
\ifarxiv\else
\graphicspath{{Figures/}}
\fi

\usepackage{enumerate}
\usepackage{mwe}
\usepackage{caption}
\usepackage{xcolor}
\usepackage[ruled,vlined,linesnumbered]{algorithm2e}
\usepackage[noend]{algpseudocode}

\newlength\myindent
\usepackage{amsmath}

\usepackage{amssymb}
\usepackage{complexity}
\usepackage{graphicx}
\usepackage[export]{adjustbox}
\usepackage{thmtools}
\usepackage{amsthm}
\usepackage{thm-restate}
\usepackage{xspace}

\usepackage[colorlinks,linkcolor=blue,citecolor=blue,filecolor=blue,urlcolor=blue]{hyperref}

\newtheorem{oq}{Open Question}

\ifprocs
\newtheorem{theorem}[lemma]{Theorem}

\newtheorem{corollary}[lemma]{Corollary}
\newtheorem{observation}[lemma]{Observation}
\newtheorem{claim}[lemma]{Claim}
\newtheorem{definition}[lemma]{Definition}
\newtheorem{hypothesis}[lemma]{Hypothesis}
\newtheorem{proposition}[lemma]{Proposition}

\else
\newtheorem{theorem}{Theorem}[section]
\newtheorem{lemma}[theorem]{Lemma}

\newtheorem{definition}[theorem]{Definition}

\newtheorem{openq}[theorem]{Open Question}

\newtheorem{corollary}[theorem]{Corollary}
\newtheorem{hypothesis}[theorem]{Hypothesis}
\newtheorem{proposition}[theorem]{Proposition}

\theoremstyle{plain}
\newtheorem{claim}[theorem]{Claim}

\fi

\ifprocs
\else

\makeatletter
\newtheorem*{rep@theorem}{\rep@title}
\newcommand{\newreptheorem}[2]{%
\newenvironment{rep#1}[1]{%
 \def\rep@title{#2 \ref{##1}}%
 \begin{rep@theorem}}%
 {\end{rep@theorem}}}
\makeatother
\newreptheorem{theorem}{Theorem}

\makeatletter
\newtheorem*{rep@corollary}{\rep@title}
\newcommand{\newrepcorollary}[2]{%
\newenvironment{rep#1}[1]{%
 \def\rep@title{#2 \ref{##1}}%
 \begin{rep@corollary}}%
 {\end{rep@corollary}}}
\makeatother
\newreptheorem{corollary}{Corollary}

\newreptheorem{lemma}{Lemma}

\def\compactify{\itemsep=0pt \topsep=0pt \partopsep=0pt \parsep=0pt}

\newcommand{\colnote}[3]{\textcolor{#1}{$\ll$\textsf{#2}$\gg$}}

\ifcomments
\newcommand{\rnote}[1]{\colnote{red}{#1--Robi}{RK}}
\newcommand{\anote}[1]{\colnote{olive}{#1--Amir}{AA}}
\newcommand{\onote}[1]{\colnote{blue}{#1--Ohad}{OT}}
\else
\newcommand{\rnote}[1]{}
\newcommand{\anote}[1]{}
\newcommand{\onote}[1]{}
\fi

\newcommand{\ProblemName}[1]{\textsf{#1}}
\newcommand{\MF}{\ProblemName{Max-Flow}\xspace}

\newcommand{\GMC}{\ProblemName{Global-Min-Cut}\xspace}

\newcommand{\pC}{\textsc{Isolating-Cuts}\xspace}
\newcommand{\pLefty}{\textsc{Lefty}\xspace}
\newcommand{\pRighty}{\textsc{Righty}\xspace}

\newcommand{\APMF}{\ProblemName{All-Pairs Max-Flow}\xspace}

\newcommand{\APSP}{\ProblemName{All-Pairs Shortest-Paths}\xspace}

\newcommand{\GHT}{Gomory-Hu Tree\xspace}

\newcommand{\GHEPT}{\ProblemName{GH-Equivalent Partition Tree}\xspace}
\newcommand{\dem}{\mathbf{d}\xspace}

\DeclareMathOperator{\cdeg}{cdeg} 
\DeclareMathOperator{\size}{size} 
\DeclareMathOperator{\vol}{vol} 

\let\poly\relax
\DeclareMathOperator{\poly}{poly}

\newcommand\eps{\varepsilon}
\renewcommand\epsilon{\varepsilon}
\newcommand\tO{\ensuremath{\tilde O}}

\newcommand{\calA}{\mathcal{A}}
\newcommand{\calY}{\mathcal{Y}}

\newcommand{\T}{\mathcal{T}}
\newcommand{\TG}{\mathcal{T}^*}
\newcommand{\TGP}{\mathcal{T}^*_p}

\newcommand{\TGPW}{\mathcal{T}^{*,\geq w}_p}

\providecommand{\set}[1]{{\{#1\}}}
\providecommand{\card}[1]{\lvert#1\rvert}

\begin{document}

\ifprocs

\title{A New GH Algorithm\thanks{A full version appears at \href{http://arxiv.org/abs/---}{arXiv:---}}}

\else


\title{APMF < APSP? \\  Gomory-Hu Tree for Unweighted Graphs \\ in Almost-Quadratic Time}

%




\author[1]{Amir Abboud\thanks{\texttt{amir.abboud@weizmann.ac.il}}}
\author[1]{Robert Krauthgamer\thanks{\texttt{robert.krauthgamer@weizmann.ac.il}}}
\author[1]{Ohad Trabelsi\thanks{\texttt{ohad.trabelsi@weizmann.ac.il}}}
\affil[1]{Weizmann Institute of Science}

\date{}

\fi

\maketitle

\ifprocs
\fancyfoot[R]{\scriptsize{Copyright \textcopyright\ 2020 by SIAM\\
Unauthorized reproduction of this article is prohibited}}
\fi 

 \thispagestyle{empty}
 \setcounter{page}{0}

\begin{abstract}
\ifprocs
\small
\fi

We design an $n^{2+o(1)}$-time algorithm that constructs a cut-equivalent (Gomory-Hu) tree of a simple graph on $n$ nodes.
This bound is almost-optimal in terms of $n$, and it improves on the recent $\tilde{O}(n^{2.5})$ bound by the authors (STOC 2021), which was the first to break the cubic barrier. 
Consequently, the All-Pairs Maximum-Flow (APMF) problem has time complexity $n^{2+o(1)}$, and for the first time in history, this problem can be solved faster than All-Pairs Shortest Paths (APSP). 
We further observe that an almost-linear time algorithm (in terms of the number of edges $m$) is not possible without first obtaining a subcubic algorithm for multigraphs.

Finally, we derandomize our algorithm, obtaining the first subcubic deterministic algorithm for Gomory-Hu Tree in simple graphs, showing that randomness is not necessary for beating the $n-1$ times max-flow bound from 1961.
The upper bound is $\tilde{O}(n^{2\frac{2}{3}})$ and it would improve to $n^{2+o(1)}$ if there is a deterministic single-pair maximum-flow algorithm that is almost-linear.
The key novelty is in using a ``dynamic pivot'' technique instead of the randomized pivot selection that was central in recent works.
\end{abstract}


\newpage

\section{Introduction}

Connectivities (minimum-cut or maximum-flow) and distances (or shortest path) are perhaps the two most fundamental measures in graphs. 
Their computational complexity is a central object of study in algorithms and discrete optimization, and both have been extensively investigated in almost any setting of interest.
Researchers often ponder the question: at a high-level, which of the two is harder?

The focus of this paper is on simple graphs (undirected, unweighted, no parallel edges or self-loops),
denoting the input graph by $G$ and its number of nodes by $n=|V(G)|$. 
For a single-pair $s,t \in V(G)$ both measures can be computed in $\tilde{O}(n^2)$ time, albeit using very different algorithms.
For shortest path, Dijkstra's algorithm~\cite{Dijkstra59} from 1956 is sufficient, while either continuous optimization \cite{linearflow21} or randomized contractions \cite{KL15} are needed for maximum-flow.
For the more demanding task of computing these measures for \emph{all-pairs} in a given graph, the complexities appear to differ.
Seidel's algorithm \cite{seidel1995all} solves all-pairs shortest paths in time $n^{\omega+o(1)}$, where $\omega\leq 2.37286$ is the fast matrix multiplication exponent \cite{AlmanW20}; whether faster algorithms are possible is one of the most well-known open questions in Algorithms.
The key breakthrough for all-pairs maximum-flow came in 1961 with the fundamental discovery of Gomory and Hu~\cite{GH61} that any graph can be turned into a tree while preserving the minimum cuts for all pairs. 
Once the tree is obtained, all pairwise connectivities can be extracted in $\tilde{O}(n^2)$ time.

\begin{theorem}[Gomory and Hu~\cite{GH61}]
Every undirected graph $G$ (even with edge weights)
has an edge-weighted tree $T$ on the same set of vertices $V(G)$ such that:
\begin{itemize} \compactify 
\item for all pairs $s,t\in V(G)$ the minimum $(s,t)$-cut in $T$ is also a minimum $(s,t)$-cut in $G$, and their values are the same.
\end{itemize}
Such a tree is called a \emph{cut-equivalent tree}, aka \emph{\GHT}. 
Moreover, the tree can be constructed in the time of $n-1$ invocations to a \MF algorithm.%
\footnote{The notation \MF refers to the maximum $(s,t)$-flow problem, 
  which clearly has the same value as minimum $(s,t)$-cut.
  In fact, we often need algorithms that find an optimal cut (not only its value),
  which is clearly different (and usually harder) than \GMC.
}
\end{theorem}

Such a strong structural result is not possible for shortest paths (e.g., because any compression to $\tilde{O}(n)$ bits must incur large error \cite{TZ05,AB16}).
Still, up until now, it has not lead to an algorithm solving the all-pairs maximum-flow problem (denoted \APMF) faster than all-pairs shortest path (\APSP).
The original algorithm of Gomory and Hu for getting such a tree has time complexity $\Omega(n^3)$, and only very recently a subcubic $\tilde{O}(n^{2.5})$-time algorithm was found \cite{AKT21}.
The immediate open question is whether the bound can be pushed all the way down to $n^2$, or perhaps the tools of fine-grained complexity could come in the way and establish a conditional lower bound.

\begin{openq}
\label{oq1}
Can one construct a \GHT of a simple graph $G$ and solve all-pairs maximum-flow in $\tilde{O}(n^2)$-time?
\end{openq}

Positive answers were obtained recently \cite{AKT20_b,LP21}, but only for $(1+\eps)$-approximate cuts.
On the negative side, the Strong Exponential Time Hypothesis (SETH) gives an $n^{3-o(1)}$ lower bound for the harder setting of \emph{directed graphs} \cite{KT18} (see also \cite{A+18} for a higher lower bound), but probably not for undirected graphs \cite{AKT20}.
Recent work has identified a class of problems that are conjectured to have $n^{2.5}$ complexity \cite{LPV20} (including all-pairs shortest paths in directed graphs \cite{CVX21}); could our problem be one of them?

\paragraph{Main Result}

For unweighted simple graphs, our main result resolves (up to $n^{o(1)}$ factors) the complexity of all-pairs maximum-flow and of the \GHT problem in dense graphs, where $\Omega(n^2)$ is a lower bound due to the input size (and for the former problem also output size). 

\begin{theorem}
\label{thm1}
There is a randomized algorithm, with success probability $1-1/\poly(n)$, that 
constructs a \GHT of a simple graph $G$ and solves \APMF in time $n^{2+o(1)}$.
\end{theorem}

We find this result surprising for multiple reasons.
First, it shows that all $n^2$ answers can be computed in the same time,
up to lower-order factors, as it takes to compute a single-pair maximum-flow.
Second, for the first time in history, the time complexity of \APMF goes below that of \APSP. 
This might be counter-intuitive because of the apparent unruliness of flows compared to paths,
e.g., the $n^{2+o(1)}$-time single-source algorithm was much more difficult to obtain.
Perhaps the only indication for the plausibility of this outcome was given in our previous paper,
where it was shown that truly subcubic time is possible for \APMF by only using combinatorial methods,
whereas it is conjectured to be impossible for \APSP \cite{AKT21}.
Third, the algorithm not only solves \APMF but also produces a \GHT that, among other things, is a space-optimal data structure for answering minimum cut invocations in $\tilde{O}(1)$ time.
While the Gomory-Hu algorithm reduces the problem to $n-1$ \MF computations, 
our new algorithm can be viewed as a reduction to only $\tilde{O}(1)$ computations on graphs with $n^{2+o(1)}$ edges (but possibly more than $n$ nodes).%
\footnote{However, viewing it this way is not sufficient for getting an almost-quadratic algorithm because the current \MF algorithms are not linear-time for all edge densities.
  Our algorithm actually uses a subroutine due to~\cite{BHKP07} that cannot be reduced to \MF computations.
}

\paragraph{Gomory-Hu Tree in $\tilde{O}(m)$ Time?}
Looking ahead,
after achieving the optimal exponent for the number of nodes $n$,
the next outstanding question is a bound in terms of the number of edges $m=|E(G)|$:
Is there an almost-linear time algorithm? 
Such a bound is known for planar graphs~\cite{BSW15}, surface-embedded graphs~\cite{BENW16}, and bounded-treewidth graphs~\cite{ACZ98,AKT20_b}.
It is also known in simple graphs if the algorithm is allowed to make nondeterministic guesses \cite{AKT20}.

\begin{openq}
\label{oq2}
Can one construct a \GHT of a simple graph $G$ in $m^{1+o(1)}$-time?
\end{openq}

One issue when studying this question is that even single-pair \MF is not known to be in almost-linear time when $m=O(n^{1.5-\eps})$. If our goal is to understand the complexity of constructing a \GHT, it is natural (and common) to assume the following plausible hypothesis.

\begin{hypothesis}
\label{hypo1}
Single-Pair \MF on $m$-edge weighted graphs can be solved in time $m^{1+o(1)}$.
\end{hypothesis}

This hypothesis does resolve Open Question~\ref{oq2} positively, even for general (weighted) graphs, if we are only interested in a $(1+\eps)$-approximate \GHT \cite{LP21}, but not in the exact case.
In the regime of sparse simple graphs, $m=O(n)$, even under this hypothesis,
the fastest algorithm for \GHT \cite{AKT20} has complexity $\tilde{O}(n^{1.5})$, leaving a gap of about $\sqrt{n}$.
For general $m$, the state of the art would be $(\min(n^2,m^{1.5}))^{1+o(1)}$,
due to Theorem~\ref{thm1} and~\cite{AKT20}. 

Our next observation is a reduction showing that
improving the $n^{1.5}$-bound for simple graphs requires
a breakthrough subcubic algorithm for \GHT in more general settings. 
However, the recent advances in algorithms for \GHT have been insufficient for breaking the cubic barrier even in unweighted non-simple graphs (\emph{multigraphs}),
which seem to be the main step towards weighted graphs;
indeed, the known bound is $\tO(mn)$ \cite{BHKP07,KL15},
which is $\tO(n^3)$ in the worst-case.

\begin{hypothesis}
\label{nonsimple}
No algorithm can construct a \GHT of an unweighted multigraph with $n$ nodes and $O(n^2)$ possibly parallel edges in time $O(n^{3-\eps})$ for a fixed $\eps>0$.
\end{hypothesis}

\begin{theorem}
\label{thm:reduction}
Assuming Hypothesis~\ref{nonsimple},
no algorithm can construct a \GHT of a simple graph on $n$ nodes and $m=O(n)$ edges in time $n^{1.5-\eps}$ for a fixed $\eps>0$.
\end{theorem}

\paragraph{Derandomization}
Taking a step back, we ask whether the developments in recent years \cite{BHKP07,AKT20,AKT20_b,AKT21} can give faster \emph{deterministic} algorithms for \GHT.
In fact, no deterministic algorithm (in any density regime) faster than Gomory-Hu's deterministic $n \cdot MF(n,m) = \Omega(nm)$ algorithm is known.
In fact, to our knowledge, the best upper bound in terms of $n$ is $O(n^{3\frac{2}{3}})$ using Goldberg and Rao's deterministic \MF algorithm \cite{GR98}.
The main difficulty is due to the fact that all these new algorithms solve the \emph{single-source} minimum-cut problem in some clever way, and then use a \emph{randomized pivot} technique, in which a uniformly random source serves as a pivot in a recursive process of logarithmic depth.
The algorithm in Theorem~\ref{thm1} (and in~\cite{AKT21})
uses randomization also in other steps,
such as expander decompositions and randomized hitting sets,
but these can already be derandomized with existing methods. 

A central contribution of our work is a new \emph{dynamic pivot} technique that can (sometimes) derandomize the randomized-pivot technique at no extra cost.%
The main result in this context is the first subcubic algorithm for \GHT.
 
\begin{theorem}
\label{thm2}
There is a deterministic algorithm that constructs a \GHT of a simple graph
and solves \APMF in time $\tilde{O}(n^{2\frac{2}{3}})$. 
The time bound improves to $n^{2+o(1)}$ assuming that single-pair \MF can be computed in deterministic time $m^{1+o(1)}$ in $m$-edge weighted graphs.
\end{theorem}



\subsection{Prior Work}
The literature on cut-equivalent (Gomory-Hu) trees is quite vast. 
They have found many important applications in several fields (see \cite[Section 1.4]{AKT21}), and have been generalized in several ways.
For example, they can be adapted to support other cut requirements such as multiway cuts or cuts between groups of nodes \cite{Hassin88,Hassin90,Hassin91,Har01,EH05,CKK16}.
The possibility of such magical representations of the minimum $s,t$-cuts in a graph have also been investigated in the harder settings of vertex connectivity and directed graphs (see \cite{Benczur95,HL07,PY21} and the references therein).

\paragraph{Previous Algorithms for \GHT}
Over the years, the time complexity of constructing a \GHT has decreased several times due to improvements in \MF algorithms, but there have also been conceptually new algorithms. 
Gusfield~\cite{Gusfield90} presented a modification of the Gomory--Hu algorithm in which all the $n-1$ calls to \MF
are made on the original graph $G$ (instead of on contracted graphs). 
Bhalgat, Hariharan, Kavitha, and Panigrahi~\cite{BHKP07} designed an  $\tilde{O}(mn)$-time algorithm utilizing a tree packing approach~\cite{Gabow95,Edmonds70} that has also been used in other algorithms for cut-equivalent trees~\cite{Cole03,HKP07,AKT20,AKT21}.
In particular, they designed an $O(mk)$-time algorithm for constructing a $k$-partial \GHT, which preserves the minimum cuts if their size is up to $k$ (see \cite{Panigrahi16} and the full version \cite{BCHKP08}).
A simple high-degree/low-degree strategy is helpful in sparse graphs:
only $\sqrt{m}$ nodes can have degree (and therefore outgoing flow) above $\sqrt{m}$, thus a $\sqrt{m}$-partial tree plus $\sqrt{m}$ \MF queries are sufficient,
which takes $O(m^{3/2})$ time if \MF is solved in linear time.
Using the current \MF algorithms, this strategy results in the time bound $\tO(\min\set{m^{3/2} n^{1/6}, \max\set{mn^{3/4},m^{3/2}}})$ \cite{AKT20}. 
Combined with \cite{AKT21}, the state of the art before this work was 
$\tO(\min\set{m^{3/2} n^{1/6}, mn^{3/4}, n^{3/2}m^{1/2}})$ 
and now it is 
$\min(n^{2+o(1)},\tO(m^{3/2} n^{1/6}),\tO(mn^{3/4}))$.
Additionally, two recent algorithms accelerate the Gomory--Hu algorithm by making multiple steps at once to achieve $\tilde{O}(m)$ time, one requires nondeterminism \cite{AKT20} and the other requires a (currently non-existent) fast minimum-cut data structure \cite{AKT20_b}.

In weighted graphs, the bound for \GHT is still $n$ times \MF and therefore cubic using the very recent $\tilde{O}(m+n^{1.5})$ algorithm for \MF \cite{linearflow21}.
If the graph is sparse enough, one can use the recent $\tO(m^{3/2-1/328})$ time algorithm~\cite{GaoSparse21}. If additionally the largest weight $U$ is small, a series of algorithms exist~\cite{madry2016computing,LS19,LS20FOCS}
that run in time $\tilde{O}( \min\{m^{10/7}U^{1/7}, m^{11/8}U^{1/4},$ $m^{4/3}U^{1/3}\} )$
and might give a better time bound.
The main open question left by this work is whether the ideas behind the drastic speed ups for unweighted graphs will lead to faster algorithms for weighted graphs as well.

Practice-oriented algorithms and experimental studies of some algorithms for \GHT can be found in \cite{akiba2016cut,GT01}.
See the Encyclopedia of Algorithms entry~\cite{Panigrahi16} for a further discussion on \GHT algorithms.

\paragraph{Approximations} The challenge of designing approximation algorithms for \GHT has received considerable attention lately~\cite{AKT20_b,LP21}.
In what may be the final answer, Li and Panigrahy~\cite{LP21} show that an $m^{1+o(1)}$ time algorithm constructing a $(1+\eps)$-approximate \GHT, even for weighted graphs, would follow from Hypothesis~\ref{hypo1}.
 Notably, the key tool in their result is a simple yet powerful \pC procedure introduced in order to get a deterministic algorithm for the global minimum cut problem \cite{LP20} and is also crucial in our exact algorithms for \GHT in unweighted graphs \cite{AKT21} (and this work).
Interestingly, the expanders-based techniques that are also crucial in our exact algorithms were not needed for their $(1+\eps)$-approximation.

\paragraph{Expander Decompositions}
A key ingredient of our \GHT algorithms (in this paper and in \cite{AKT21}) is an expander-decomposition of the graph \cite{KVV04,OV11,OSV12,ST13,SW19,CGLNPS20,GHTZ21}.
Such decompositions have led to several breakthroughs in algorithms for basic problems in the past decade, e.g. \cite{ST14,KLOS14,NSW17}.
Even more recently, expander decompositions and conductance based arguments have been used to break longstanding bounds for other problems related to minimum cuts such as: deterministic algorithms for global minimum cut \cite{KThorup19,saranurak2020simple,LP20,Li21} and vertex connectivity \cite{Gabow06,NSY19,FNYSY20}.

The added difficulty of the \GHT problem compared to the global minimum cut problem comes from the low degree nodes.
When we are searching for a minimum $s,t$-cut of value $w$ there could be many nodes with degree $<w$ that must be assigned to the correct side of the cut.
But in the global minimum cut problem this cannot happen: if the global minimum cut has value $w$ then all nodes in the graph must have degree $\geq w$.
These low-degree nodes create difficulties for the expansion-based arguments because, intuitively, they do not contribute as much as expected to the volume (for a more detailed explanation see \cite[Section 1.3, Complication 5]{AKT21}).


%
%
%
%
%


\subsection{Technical Overview}
\label{overview}

In this section we distill and highlight the novel technical ideas of this work,
instead of providing an overview of the actual algorithms. 
The overview below is a considerable over-simplification,
because the algorithms in both the work we improve upon~\cite{AKT21} and (to a lesser extent) this paper combine intricately many ingredients, and they cannot be truly explained without many preliminaries that may only be known to experts.
Nevertheless, the standard/advanced terminology and algorithms mentioned below
without explanation can all be found in the Preliminaries (Sections~\ref{sec:prelims} and~\ref{sec:prelimsDet}). 

The plan is to first motivate and explain the three new ideas that are the key to the new results. 
We hope that they can be appreciated on their own even if it is not possible to see exactly how they fit into the actual algorithm.
Afterwards, we will discuss some of the complications that arise when using these ideas inside the algorithm.

\subsubsection{The Simplest Hard Tree: Breaking the $n^{2.5}$ Barrier using Expander-Decomposition with Demands}

\begin{figure}[ht]
  \begin{center}
    \includegraphics[width=2.3in]{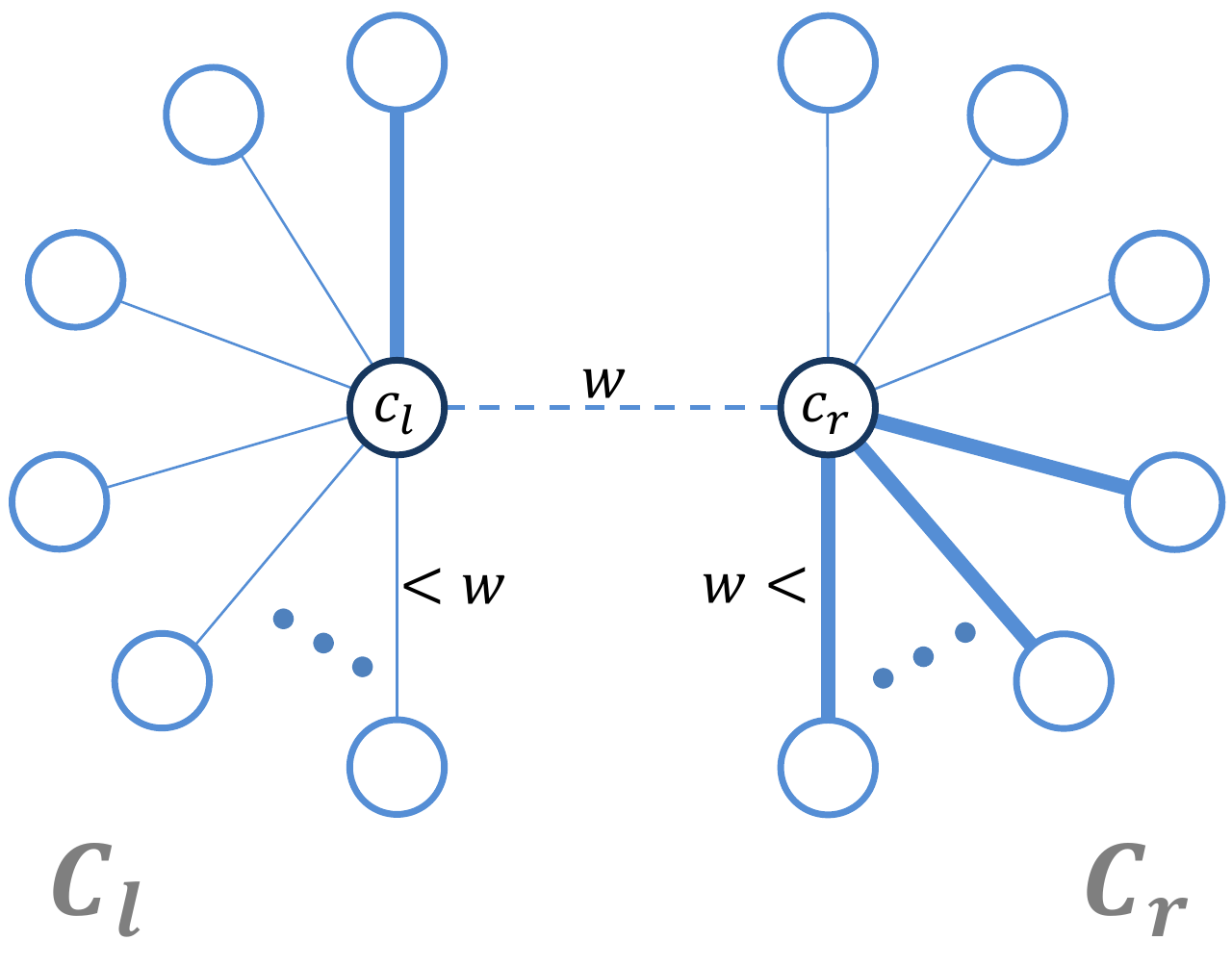}
    \end{center}
    \caption{The cut-equivalent tree $T$ of a hard case, where $c_r$ is given and the goal is to identify $c_\ell$, the only node with a non-trivial minimum cut to $c_r$.
      The weight of the cut $(C_\ell,C_r)$ is $\lambda_{c_l,c_r} = w$, illustrated by the dashed edge at the center; edges of weight $>w$ are thick, and edges of weight $<w$ are thin.
      The minimum cut between any pair of leaves $u,v$ is trivial unless $u$ is incident to a bold edge on the left and $v$ is incident to a bold edge on the right.
    }
    \label{Figs:clcr}
\end{figure}

The simplest hard case for constructing a \GHT appears to be the following.\footnote{It had appeared to be hard both for breaking the $n^{2.5}$ bound in simple graphs and, until today, for breaking the $n^3$ bound in multigraphs and weighted graphs. This is discussed in Section 3 in \cite{AKT21}.}
Given a graph $G$ distinguish between the case where its only \GHT is a star from the case where it is a $c_{\ell}$-$c_{r}$ tree; i.e. two stars connected by an edge between the two centers $c_{\ell}$ and $c_r$.
In the first case, the graph has the property that the minimum $u,v$-cut for all nodes $u,v \in V$ is one of the two trivial cuts $\{u\}$ or $\{v\}$,
and in the second case there is a single non-trivial minimum cut $(C_\ell,C_r)$, the cut between $c_{\ell}$ and $c_r$, while almost all others are trivial.
This is a hard case even if we are given the node $c_r$ in advance; it is helpful to think of it as the pivot or source from which we want to compute single-source cuts.
The difficulty is in finding $c_{\ell}$ (if it exists at all).
To make it more concrete (and still hard) suppose that both stars have $\Omega(n)$ nodes and the minimum $c_{\ell},c_{r}$-cut has value $w = \Theta(n)$ which in turn implies that both $c_{\ell},c_{r}$ must have degree $\Omega(n)$.  
See Figure~\ref{Figs:clcr}.

It is instructive to observe that random sampling does not help, even if the cut we are searching for (the minimum $c_{\ell},c_r$-cut) is balanced.
Even if we sample a node $\ell$ from $c_{\ell}$'s star and a node $r$ from $c_r$'s star (or $c_r$ itself), their minimum $\ell,r$-cut will not reveal the cut we are searching for; it could even be simply $\{\ell\}$ or $\{r\}$.\footnote{This is a big difference from related problems that ask for the global connectivity of a graph. There, finding a node from each side is the end of the story.}
It seems that $\Omega(n)$ calls to a \MF algorithm are required with this strategy.
In fact, random sampling is one of the ways that help one see that $c_{\ell}$-$c_{r}$ is the ``hardest tree'': if the tree had a long path it would have been possible to shorten it with a randomized hitting set.
Another method, the powerful \pC procedure \cite{LP20}, lets one resolve trees with many small subtrees, but it too fails against the $c_{\ell}$-$c_{r}$ tree where there is only one large subtree (see \cite[Section 3]{AKT21} for more details).
One important observation, that is the first step towards a solution is that $c_{\ell}$ is the highest degree node among the nodes in the left star $C_{\ell}$;\footnote{Suppose for contradiction that one of the leaves $v \in C_{\ell}$ has higher degree than $c_{\ell}$. Then $\{c_{\ell}\}$ is a better cut than $\{v\}$ the minimum cut in the tree, violating the requirements of a \GHT.} but the right star could have many nodes with higher or lower degree than it.

Perhaps the main idea in the subcubic algorithm of \cite{AKT21} was to use an expander decomposition of $G$ in order to locate the node $c_{\ell}$ using fewer or cheaper calls to a \MF algorithm.
The graph is decomposed into $\phi$-expanders $H_1,\ldots,H_h$ with expansion parameter $\phi = 1/\sqrt{n}$ and the number of outer-edges (that leave an expander) is $\tilde{O}(|E|\phi)=\tilde{O}(n^{1.5})$.
Since the cut we are searching for has only $O(n)$ edges,
no expander can contain $\Omega(\sqrt{n})$ nodes from each side of this cut,
as that would violate the $\phi$-expansion requirement,
as it would induce inside the expander a cut of conductance
$\frac{O(n)}{\Omega(\sqrt{n})} < \phi$
(see Section~\ref{sec:exp-decomp} for relevant definitions). 
Now, a key observation that crucially relies on the fact that $G$ is a simple graph, is that only $\tilde{O}(\sqrt{n})$ nodes of high degree, say at least $n/10$, can be in small expanders $H_i$, i.e., of size at most $n/100$;
this is because such a high degree node in a small expander must have at least $n/10-n/100 \ge \Omega(n)$ incident outer-edge,
and there can only be $\tilde{O}(n^{1.5})$ outer-edges in the decomposition.
On the other hand, there can only be $O(1)$ large (i.e., not small) expanders
since there are only $n$ nodes in the graph.
Consequently, it is enough to search for $c_{\ell}$ inside the $O(1)$ large expanders, using the clever methods that follow, or among those $\tilde{O}(\sqrt{n})$ high-degree nodes using direct calls to a \MF algorithm.
To search in large expanders two methods are used: one is useful when there are few nodes from the left star in the expander, 
and one is useful when there are few nodes from the right star.
(Recall that no expander can contain $\Omega(\sqrt{n})$ nodes from each star.)

In brief, the first procedure uses the fact that the expander contains $c_{\ell}$ plus only few nodes from $C_{\ell}$ in order to find a set of nodes $S$ that isolates $c_{\ell}$, in the sense that $C_{\ell} \cap S = \{c_{\ell} \}$, and then uses the \pC procedure.
And the second procedure, which will be elaborated on in the next subsection, uses the fact that $c_{\ell}$ is the highest degree node in $C_{\ell}$ to conclude that it must be among the $O(\sqrt{n})$ highest-degree nodes in an expander that has $<\sqrt{n}$ nodes from $C_r$.
The upshot is that $\tilde{O}(\sqrt{n})$ calls to a \MF algorithm are enough to resolve any expander, and since this is only done for the $O(1)$ large expanders, the $n^{2.5}$ bound follows.

How can we beat this $\sqrt{n}$ barrier? 
It is natural to make the expansion parameter $\phi$ larger, e.g. $1/\log^{O(1)}{n}$ rather than $1/\sqrt{n}$, so that each expander is guaranteed to have fewer nodes from one of the sides, e.g., $\log^{O(1)}{n}$ rather than $\sqrt{n}$, and then it can be handled more efficiently.
The issue is that the upper bound on the number of outer-edges $\tilde{O}(\phi |E|)$ becomes worse and we end up with $\Omega(n/\log^{O(1)}{n})$ high-degree nodes in small expanders that must be checked.
On the other hand, making $\phi$ smaller, e.g. $1/n$, makes the time of handling an expander correspond to $\Omega(n)$ \MF calls.
Thus, $\sqrt{n}$ seems like a natural sweet-spot.

The new observation is that even though $\phi=1/\log^{O(1)}{n}$ may not allow us to directly find $c_{\ell}$ (since it may be among the $\Omega(n/\log^{O(1)}{n})$ high-degree nodes in small expanders), it does efficiently reduce the set of candidate for being $c_{\ell}$ from $\Omega(n)$ to about $n/\log^{O(1)}{n}$.
If this candidate elimination can be repeated, we will quickly, after $\tilde{O}(1)$ times, reach a set of only $\tilde{O}(1)$ candidates that can be checked directly.
To make this repetition possible and to avoid pointlessly eliminating the same set of candidates each time, we utilize a more powerful expander-decomposition with \emph{vertex demands},
that lets us set to $0$ the demand of nodes we have already handled.
These demands control much better how the vertices of a single expander
are split between the two stars,
because we can count only nodes that have non-zero demand and are not yet handled.
A similar strategy was used by Li and Panigrahy~\cite{LP20} for their deterministic global min-cut algorithm (in weighted graphs), a related but very different problem.
The two algorithms share a certain high-level strategy,
even though almost all the details are different.
These two striking applications of expander-decomposition with vertex demands
indicate that we are likely to see additional applications of this powerful technology.

\subsubsection{From One Hard Cut to Single-Source Cuts using New Structural Results}

After resolving the hard case of the previous section,
where there was only one ``candidate'' $c_{\ell}$ with a difficult-to-find cut,
we move on to the more general setting.
Suppose we are given a source (or pivot) node $p$, such as $c_r$ in the previous example, and want to compute the minimum cut to all other nodes $v \in V$.
For concreteness and simplicity, assume that we are only interested in cuts of value between $w$ and $2w$, where $w = \Theta(n)$.
It might seem that the strategy of the previous section simply works because $c_{\ell}$ could have been any high-degree node: we take all nodes of degree $\geq w$ to be candidates and the $\tilde{O}(1)$ calls to \MF are magically supposed to find the minimum $p,v$-cut for all of them. 

Alas, from a closer look we see that the fact that $c_{\ell}$ is the highest degree node in its cut $C_{\ell}$ is crucially used in the second of the two methods described above, and this will not be the case for all high-degree nodes; from now on we will call this second method Procedure \pLefty because it handles expanders with few nodes from the right side. 
The strategy taken in this paper and also (in a much less efficient way) in~\cite{AKT21} is to talk about \emph{estimates} instead of degrees.
The estimate $c'(v)$ of a node $v$ is initially the degree of $v$,
and it is reduced whenever a $p,v$-cut of smaller value is found. Eventually, when a node is ``done'' the estimate becomes equal to the value of the minimum $p,v$-cut (that is always upper bounded by the degree of $v$).
In Procedure \pLefty in \cite{AKT21},
instead of taking the $O(\phi^{-1})$ highest degree nodes from an expander, we actually take the highest estimate nodes.
This lets us compute correct cuts not only for nodes that have the highest degree among their cut-members, but also those whose current estimate is highest.
If this (the entire algorithm) is repeated enough times, letting the estimates improve after each repetition, it is guaranteed that all the minimum cuts will be found.\footnote{To get intuition, observe that if a node $v$ is the only node in its cut $C_v$ that is not done, then it must have a higher estimate than all other nodes in $C_v$. The same cannot be said about the degree.}

But how many times should the algorithm be repeated?
In \cite{AKT21} it is argued that $\tilde{O}(1)$ repetitions are sufficient because the algorithm has the budget to make $\sqrt{n}$ additional invocations each time Procedure \pLefty is called and because the depth of the \GHT can be upper bounded by $\sqrt{n}$ by taking a random hitting set of $\sqrt{n}$ nodes.
Unfortunately, both of these things are not possible if we want $n^{2+o(1)}$ time.
First, we cannot spend $\sqrt{n}$ additional invocations and second, the depth reduction itself already has $\Omega(n^{2.5})$ running time.

The approach taken in this paper is different, much more efficient, and exploits the fact that the graph is simple in a deeper way.
We do not repeat the algorithm at all, but we augment Procedure \pLefty so that it does not stop handling an expander until the node with highest estimate in the expander is done.
The number of calls to \MF can no longer be upper bounded by $O(\phi^{-1})=\tilde{O}(1)$ and, in principle, some expanders might incur a much larger number of calls.
But we can provide an upper bound on the total number of calls using the fact that each additional call was a result of a previously undone node becoming done, meaning that a new minimum cut was found. In fact, the new cut is also not ``easy'' in the sense that it contains at least two high-degree nodes; this important point will not be discussed in this overview (except briefly in Section~\ref{sec:complications} Item~\ref{complication_easy}).
Then, using a new bound on a certain notion of depth of a \GHT of a simple graph, we can upper bound the number of such cuts by $O(n/w)$ which is a negligible overhead for our algorithm.
This bound is proved in Section~\ref{sec:structural} using combinatorial arguments of the following flavor:
Suppose there is a set of $k$ minimum cuts of value between $w$ and $2w$ that nested (each cut is a subset of the previous one). 
If most of them have ``private sets'' of $\Omega(w)$ nodes that do not belong to the other cuts, then $k=O(n/w)$, as desired. 
Otherwise, there is a sequence of $10$ cuts with only $o(w)$ private nodes each; now, because the graph is simple and each of these cuts must contain a node of degree $\geq w$, we get that each of these cuts has $(1-o(1)) w$ edges leaving the entire set of $10$ cuts.
But then there are $(10-o(1)) w$ edges leaving the largest of these cuts,
contradicting the fact that its value is $\leq 2w$.
This structural result follows from basic combinatorics, 
but when used properly (e.g., one has to bypass complications with easy cuts), 
it can lead to significant algorithmic savings.

\subsubsection{Derandomization using a Dynamic Pivot}

Until now, we have described how to find the minimum cuts from a single pivot $p$ to all other nodes $v \in V$.
Let use remark that essentially all techniques needed for our $n^{2+o(1)}$ time single-source minimum cuts algorithm can be derandomized with standard tools.\footnote{The only exception is the $k$-partial tree that is only needed as long as the upper bound for \MF is not almost-linear. For the purposes of this overview, it is not needed.}
To fully construct a \GHT, this algorithm is used recursively in a manner that is similar to the Gomory-Hu algorithm.
There is an intermediate tree $T$ that starts from a single super-node containing all of $V$, and gets refined throughout the recursion until eventually each node is in its own super-node.
A refinement of the tree $T$ is performed by dividing a super-node into smaller super-nodes, based on information about some minimum cuts.
The Gomory-Hu algorithm divides the super-node into two each time, while our algorithm (since it has all the minimum cuts from a single pivot) may divide it into many super-nodes at once.\footnote{This is a crucial point, as discussed in our previous papers \cite{AKT20,AKT20_b}}
For purposes of efficiency, we want to make the recursion depth logarithmic.
This is guaranteed if the cuts from the pivot happen to divide the graph in a balanced way; but unfortunately, it could be the case that the minimum cut from the pivot $p$ to all other nodes is $(\{p\},V\setminus\{p\})$ and we have learned very little.
A popular way to circumvent this issue (\cite{BHKP07,AKT20_b,AKT21}),
used also in our randomized algorithm,
is to pick the pivot at random and argue that with high probability,
many of the cuts will be balanced.
But it is not clear how a deterministic algorithm
can avoid depth $\Omega(n)$ due to a sequence of bad pivots.

One approach is to use the new structural results to argue that, in a simple graph, if we always pick nodes of degree $\geq w$ as pivots, the number of calls to the single-source algorithm will be $O(n/w)$.
This would be an improvement over $\Omega(n)$ calls, but it will not lead to an $n^{2+o(1)}$ bound (even if \MF can be solved in linear time).

The approach we take is both more powerful and conceptually simpler.
We modify the single-source algorithm so that it is guaranteed to return balanced cuts: if this is not possible because the pivot is bad, then the single-source algorithm may change the pivot into a better one.
A good pivot is always guaranteed to exist, e.g. the centroid $p^*$ of the \GHT has the property that for all $v \in V$, the minimum $p^*,v$-cut has $\leq n/2$ nodes in the side of $v$.

In more detail, we start from the highest degree node as the pivot $p$ and proceed with the single-source algorithm as usual.
Whenever a cut $C_q$ for node $q\in V$ is obtained, either via a call to \MF or via the \pC procedure (Lemma~\ref{lem:proc_main}), we check to see if the cut is good in the sense that the side of $q$ has $\leq n/2$ nodes.
(We work with so-called latest cuts, see Section~\ref{sec:latest},
so that if a good cut exists, it will be found.)
If the cut is not good, the algorithm runs a \emph{pivot-change protocol} that makes $q$ the new pivot.
In this protocol, the estimates and cuts of all nodes $v \in V$ are updated;
fortunately, due to a triangle-inequality-like property of cuts,%
\footnote{It is well-known that
  $\min(\lambda_{x,y}, \lambda_{y,z}) \leq \lambda_{x,z}$ for all nodes $x,y,z$, 
  where $\lambda_{u,v}$ denotes the minimum $u,v$-cut value. 
}
it can all be done in $O(1)$ time per node $v \in V$, which is negligible compared to the time for solving the \MF leading to that pivot-change.
And that is it; the algorithm can proceed as if $q$ was the pivot all along.


\subsubsection{Further Complications}
\label{sec:complications}

Finally, let us mention the complications that arise when turning the above ideas into a full algorithm for constructing a \GHT.
Happily, some of the complications in our previous work \cite{AKT21} were alleviated due to the recent discovery of a \MF algorithm that runs in near-linear time when $m=\Omega(n^{1.5})$ \cite{linearflow21}. 

\begin{enumerate}

\item The first reason that our actual algorithms are more complicated than what has been described is that the single-source algorithm must work with auxiliary graphs of the iput graph $G$, and thees are obtained by contracting certain nodes (as in the Gomory-Hu framework).
While the input graph is simple, the auxiliary graphs could have parallel edges, and so the same arguments cannot be applied in a black-box manner.
Still, we apply them by going back and forth (in both the algorithm and analysis) between the nodes of the auxiliary graph and of the input graph.
While this does not introduce substantial technical difficulties, the notation is heavier and it takes some time to get used to.
 
\item When searching for cuts of value $w=o(n)$ the number of calls to a \MF algorithm is higher. 
Typically, our algorithms make $\tilde{O}(n/w)$ calls to a \MF algorithm.
Fortunately, as $w$ gets smaller, each call becomes cheaper because it can be applied on a Nagamochi-Ibaraki sparsifier \cite{NI92} of the graph that only has $O(nw)$ edges and preserves all cuts of value $\leq w$.
So if we had an almost-linear time \MF algorithm,
all the calls together would overall take $(nw)^{1+o(1)} \cdot n/w = n^{2+o(1)}$ time.
But using the current \MF algorithms each call takes $\tilde{O}(nw + n^{1.5})$,
which is larger when $w=o(\sqrt{n})$,
making the overall time super-quadratic.
To circumvent this issue, we start with a preliminary step that computes a $\sqrt{n}$-partial tree of the graph \cite{BHKP07} in $\tilde{O}(n^2)$ (randomized) time, and afterwards all cuts we are interested in will have $w>\sqrt{n}$.

\item In our randomized algorithm, where we use the standard randomized pivot selection to go from a single-source algorithm to a full tree, it is convenient to assume that all minimum cuts are unique.
Otherwise the \MF algorithm might adversarially return bad (unbalanced) cuts with respect to whatever pivot we choose.
This can be achieved by adding a small perturbation to each edge,
which guarantees that ties are broken in some consistent manner
and does not ruin the arguments that need the graph to be unweighted
because we can still analyze the underlying simple graph.
There are other ways to circumvent this issue; for example, by dropping the randomized pivot strategy altogether and using our dynamic pivot technique instead. Still, we have chosen to present this method because it may be the easiest to follow for readers who are familiar with prior work.

\item\label{complication_easy} Finally, there is a step in our single-source algorithm that might seem unnecessary at first.
We invoke the \pC procedure to compute all easy cuts (recall these contain only one node of degree $\geq w$).
We call them easy because they are indeed easy to compute using a single call to \MF and without any expansion-based arguments.
The need for this step is in order to bound the number of new cuts obtained in Procedure \pLefty later on in the algorithm using our new structural results.
This result can only bound the number of cuts that are not easy.
Therefore, omitting this step might make Procedure \pLefty prohibitively expensive.

\end{enumerate}



\section{Preliminaries}
\label{sec:prelims}

%
%
%
%
%

\subsection{Gomory-Hu's Algorithm and Partial Trees}
\label{prelim:GH}

\begin{figure}[ht]
  \begin{center}
    \ifarxiv
    \includegraphics[width=6.2in]{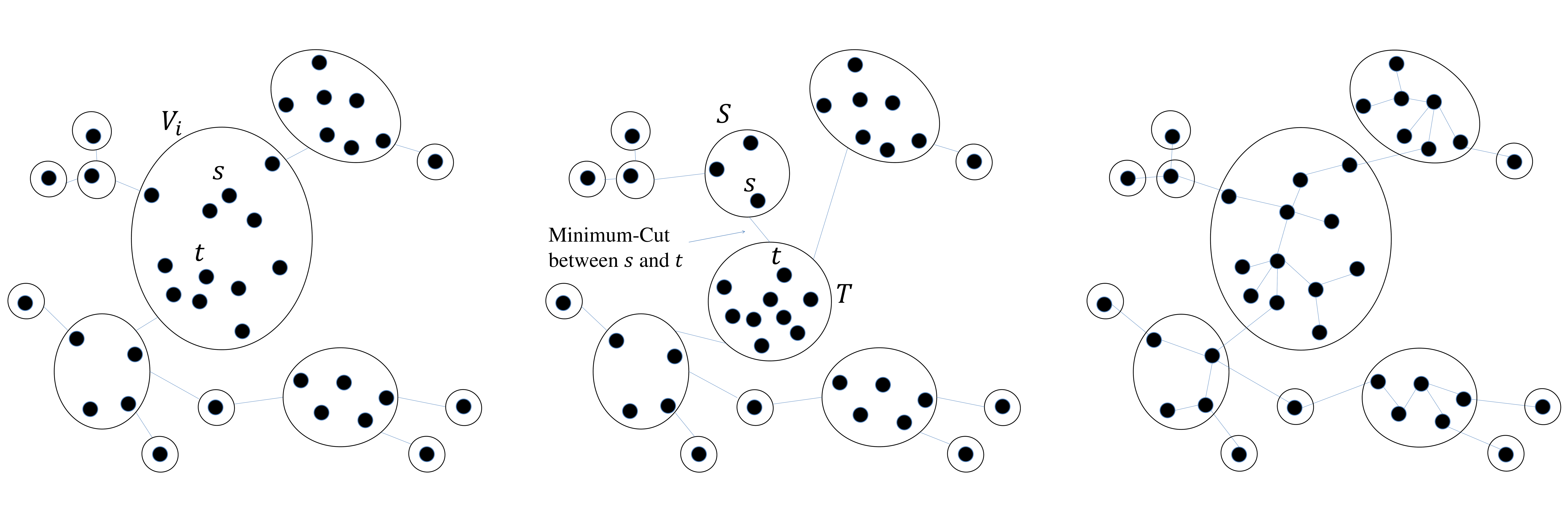}
    \else
    \includegraphics[width=6.2in]{GH_trees_3}
    \fi
    \end{center}
    \caption{An illustration of the construction of $\T$.  Left: $\T$ right before the partition of the super-node $V_i$. Middle: after the partitioning of $V_i$. Right: $\T$ as it unfolds after the Gomory-Hu algorithm finishes.}
    \label{Figs:GH_trees_3}
\end{figure}

First, we give some general definitions that are often used by algorithms for Gomory-Hu trees.
\paragraph{Partition Trees.}
A \emph{partition tree} $T$ of a graph $G=(V,E)$
is a tree whose nodes $1,\dots,l$ are \emph{super-nodes},
which means that each node $i$ is associated with a subset $V_i\subseteq V$; 
and these super-nodes form a disjoint partition $V=V_1 \sqcup\cdots\sqcup V_l$.
We will assume that the edges of a partition tree are weighted in the natural way: 
each edge in $T$ represents a vertex bipartition in $G$ and we can define its weight to be the value of this cut in $G$. 
An \emph{auxiliary graph} $G_i$ is constructed from $G$ by merging nodes that lie in the same connected component of $T\setminus \{i\}$.
These merged nodes (representing multiple nodes in $G$) will be called \emph{contracted node}. 
For example, if the partition tree is a path on super-nodes $1,\ldots,l$, 
then $G_i$ is obtained from $G$ by merging $V_1\cup\cdots\cup V_{i-1}$ into one contracted node and $V_{i+1}\cup\cdots\cup V_l$ into another contracted node. We will use the notations $n_i':=\card{V_i}$, $m_i':=\card{E(G_i)}$, and $n_i:=\card{V(G_i)}$. Note that $n_i' \leq n_i$ since $V(G_i)$ contains $V_i$ as well as some other contracted nodes, with $n_i'=n_i$ only if the tree $T$ has a single super-node.
The following is a brief description of the classical Gomory-Hu algorithm~\cite{GH61} (see Figure~\ref{Figs:GH_trees_3}).


\paragraph{The Gomory-Hu algorithm.}
This algorithm constructs a cut-equivalent tree $\T$ in iterations. 
Initially, $\T$ is a single node associated with $V$ (the node set of $G$), 
and the execution maintains the invariant that $\T$ is a partition tree of $V$.
At each iteration, the algorithm picks arbitrarily two graph nodes $s,t$ 
that lie in the same tree super-node $i$, i.e., $s,t\in V_i$.
The algorithm then constructs from $G$ the auxiliary graph $G_i$
and invokes a \MF algorithm to compute in this $G_i$ a minimum $st$-cut, denoted $C'$.
The submodularity of cuts ensures that this cut is also 
a minimum $st$-cut in the original graph $G$, 
and it clearly induces a disjoint partition $V_i=S\sqcup T$ 
with $s\in S$ and $t\in T$. The algorithm then modifies $\T$ by splitting super-node $i$
into two super-nodes, one associated with $S$ and one with $T$,
that are connected by an edge whose weight is the value of the cut $C'$,
and further connecting each neighbor of $i$ in $\T$ 
to either $S$ or $T$ (viewed as super-nodes),
depending on its side in the minimum $st$-cut $C'$
(more precisely, neighbor $j$ is connected to the side containing $V_j$).

The algorithm performs these iterations until all super-nodes are singletons, which happens after $n-1$ iteration.
Then, $\T$ is a weighted tree with effectively the same node set as $G$.
It can be shown \cite{GH61} that for every $s,t\in V$,
the minimum $st$-cut in $\T$, viewed as a bipartition of $V$,
is also a minimum $st$-cut in $G$, and of the same cut value.
We stress that this property holds regardless of the choice made at each step
of two nodes $s\neq t\in V_i$. 

A \GHEPT is a partition tree that can be obtained by a truncated execution of the Gomory-Hu algorithm, in the sense that there is a sequence of choices for the pairs $s\neq t\in V_i$ that can lead to such a tree.
A basic property of partition trees and auxiliary graphs is the following.

\begin{lemma}[\cite{GH61}]
\label{lem:auxiliary}
Let $T$ be a \GHEPT of a graph $G=(V,E)$ and let $G'$ be an auxiliary graph on original nodes $V' \subseteq V$.
For all $u,v \in V'$ it holds that $\MF_G(u,v)=\MF_{G'}(u,v)$.
\end{lemma}

The following simple lemma describes the flexibility in designing cut-equivalent tree algorithms based on the Gomory-Hu framework.

\begin{lemma}
\label{lem:partial_tree_comb}
Given a \GHEPT $T'$ of an input graph $G$, and a cut-equivalent tree $T_i$ of the auxiliary graph $G_i$ for each super-node $V_i$ in $T'$, one can construct a full cut-equivalent tree $T$ of $G$ in linear time. 
\end{lemma}

\begin{proof}
In a preprocessing step, for every super node $V_i$ in $T'$ and every contracted node $q\in  V(G_i)\setminus V_i$, save a pointer to the super node $V_j\subseteq q$ that is adjacent to $V_i$ in $T'$.
Now, for every super-node $V_i$, identify each contracted node $q\in V(G_i)\setminus V_i$ with the super-node it contains that is adjacent to $V_i$ in $T'$, and connect the nodes of $V_i$ to the super-nodes they are adjacent to (if any).
Finally, if a node $u\in V_i$ is connected to a super-node $V_j$, and a node $v\in V_j$ is connected to $V_i$, remove these connections and connect $u$ to $v$ directly, and call the result $T$. Observe that $T$ must be a tree.

To see why $T$ is a correct cut-equivalent tree of $G$, observe that there exists a simulated Gomory-Hu execution that results in $T$. Given the \GHEPT $T'$, pick pairs of nodes from $V_i$ and cuts according to $T_i$. This is guaranteed to produce a tree $\tilde{T}_i$ whose projection on $V_i$ is identical to $T_i$, while the subtrees adjacent to $V_i$ in $\tilde{T}$ are connected to the same nodes of $V_i$ as their contracted counterparts in $T_i$. 
Applying this simulated execution to all super-nodes concludes the proof.
\end{proof}

%
%
%
%
%

A simple corollary is the following natural lemma that we will use.

\begin{lemma}
\label{noncrossingGH}
Let $G$ be a graph and suppose that $S_1,\ldots,S_k$ are non-crossing minimum cuts between a single source $p$ and $k$ targets.
Then there exists a \GHT of $G$ in which all of these $k$ cuts are minimum cuts.
\end{lemma}

\begin{proof}
To construct such a tree, execute the Gomory-Hu algorithm where in the first $k$ steps choose the pairs $p,v_i$ and suppose that $S_i$ is the cut returned by the invocation. Then continue the execution arbitrarily. Since these $k$ cuts will be contracted in the later stages, the final tree will preserve them.
\end{proof}

\subsubsection{$k$-Partial Trees}
\label{prelim:kpartial}
A $k$-partial tree, formally defined below, can also be thought of as
taking a cut-equivalent tree of $G$ and contracting all edges of weight greater than $k$. 
Such a tree can obviously be constructed using the Gomory-Hu algorithm,
but as stated below (in Lemma~\ref{Lemma:Partial}), 
faster algorithms were designed in~\cite{HKP07,BHKP07},
see also~\cite[Theorem $3$]{Panigrahi16}.
It is known that such a tree is a \GHEPT, see~\cite[Lemma $2.3$]{AKT20}.

\begin{definition}[$k$-Partial Tree~\cite{HKP07}] 
  A \emph{$k$-partial tree} of a graph $G=(V,E)$ is 
  a partition tree of $G$
  with the following property:
  For every two nodes $x,y\in V$ whose minimum $x,y$-cut value in $G$ is at most $k$,
  nodes $x$ and $y$ lie in different super-nodes $x\in X$ and $y\in Y$,
  such that the minimum $X,Y$-cut in the tree defines a bipartition of $V$
  which is a minimum $x,y$-cut in $G$ and the two cuts have the same value.
\end{definition}
  
\begin{lemma}[\cite{BHKP07}]
\label{Lemma:Partial}
There is an algorithm that given an undirected graph having $n$ nodes, $m$ edges, and unit edge-capacities together with an integer $k\in [n]$, constructs a $k$-partial tree in time $\min\{\tO(nk^2),\tO(mk)\}$.
\end{lemma}

\subsection{Edge Perturbation}

The following proposition shows that by adding small capacities to the edges, we can assume that $G$ has one cut-equivalent tree $\TG$ (see also~\cite[Preliminaries]{BENW16} and~\cite{AKT20_b}).

\begin{proposition}
\label{Proposition:Perturbation}
One can add random values in $\{1/n^{10},\dots,n^7/n^{10}\}$
to the edge-capacities in $G$, such that with probability $1-1/poly(n)$, the resulting graph $\tilde{G}$ has
a single cut-equivalent tree $\TG$ with $n-1$ distinct edge weights,
and moreover the same $\TG$ (with edge weights rounded back) 
is a valid cut-equivalent tree also for $G$.
\end{proposition}

Throughout this paper, when we say that $\tilde{G}$ is a perturbed version of $G$ with unique minimum cuts we mean that $\tilde{G}$ was obtain from $G$ as in Proposition~\ref{Proposition:Perturbation}. 
The truth is that sometimes, with low probability, the perturbed version $\tilde{G}$ may not have unique minimum cuts; but this event will be ignored throughout our analysis and, if it occurs, the algorithm is assumed to fail.

\subsection{Nagamochi-Ibaraki Sparsification}

We use the sparsification method by Nagamochi and Ibaraki~\cite{NI92}.
They showed that for any graph $G$ it is possible to find a subgraph $H$ with at most $w(n-1)$ edges, such that $H$ contains all edges crossing cuts of value $w$ or less.
It follows that if a cut has value at most $w-1$ in $G$ then it has the same value in $H$, and if a cut has value at least $w$ in $G$ then it also has value at least $w$ in $H$.
Their algorithm~\cite{NI92} performs this sparsification in time $O(m)$ 
on unweighted graphs, independently of $w$, and it works for multigraphs that have $m$ parallel edges.

We will need the following simple lemma, showing that sparsification and edge-perturbation can be combined in a convenient way.

\begin{lemma}[The Nagamochi-Ibaraki sparsifier of a perturbed graph]
\label{perturbed_NI}
Given an unweighted multigraph graph $G$ on $n$ nodes and $m$ (possibly parallel) edges, an integer $w\geq 1$, and a perturbed version $\tilde{G}$ of $G$ with unique minimum cuts, one can compute in $O(m)$ time a perturbed sparsifier $G_w$ on $O(nw)$ edges such that:
\begin{itemize}
\item any cut of weight $<w$ in $\tilde{G}$ has exactly the same weight in $G_w$, and
\item any cut of weight $\geq w$ in $\tilde{G}$ still has weight $\geq w$ in $G_w$.
\end{itemize}
\end{lemma}

\begin{proof}
Let $\eps(e) \in \{1/n^{10},\dots,n^7/n^{10}\}$ be the added perturbation of each edge $e \in E(G)$ that was added in order to produce $\tilde{G}$, and note that $\eps(e)<1$.
To produce the perturbed sparsifier $G_w$ start by computing a Nagamochi-Ibaraki sparsifier $H$ of the unweighted $G$, and then for each edge $e \in E(G) \cap E(H)$ that remains in $H$, add $\eps(e)$ to its weight. 

Let $S$ be a cut of weight $<w$ in $\tilde{G}$. Since the edge perturbations are $<1$ it follows that $S$ has weight $<w$ in $G$ as well, and therefore all of its edges are preserved in the Nagamochi-Ibaraki sparsifier $H$. It follows that the weight of $S$ in $G_w$ is exactly as it is in $\tilde{G}$.

On the other hand, let $S$ be a cut of weight $\geq w$ in $\tilde{G}$.
Again, since the perturbations are so small, the weight of this cut in $G$ must have also been $\geq w$, and by the properties of the Nagamochi-Ibaraki sparsifier, also in $H$ and in $G_w$.
\end{proof}

Consequently, if we restrict our attention to the cuts of weight $< w$, the sparsifier still has unique minimum cuts.

\subsection{Expander Decomposition}
\label{sec:exp-decomp}
We mostly follow notations and definition from~\cite{SW19}.
Let $G=(V,E)$ be an undirected graph with edge capacities. Define the \emph{volume} of $C\subseteq V$ as
$\vol_G(C) := \sum_{v\in C}\cdeg_G(v)$, 
where the subscripts referring to the graph are omitted if clear from the context.
%
The \emph{conductance} of a cut $S$ in $G$ is $\Phi_G(S) := \frac{\delta(S)}{\min(\vol_G(S),\vol_G(V\setminus S))}$.
The \emph{expansion} of a graph $G$ is $\Phi_{G} := \min_{S\subset V}\Phi_G(S)$. 
If $G$ is a singleton then $\Phi_{G}:=1$ by convention. 
Let $G[S]$ be the subgraph induced by $S\subset V$, and let $G\{S\}$ denote the induced subgraph $G[S]$ but with an added self-loop $e=(v,v)$ for each edge $e'=(v,u)$ where $v\in S,u\notin S$ (where each self-loop contributes $1$ to the degree of a node), so that every node in $S$ has the
same degree as its degree in $G$.
Observe that for all $S\subset V$, $\Phi_{G[S]}\ge\Phi_{G\{S\}}$, because the self-loops increase the volumes but not the values of cuts.
We say that a graph $G$ is a \emph{$\phi$-expander} if $\Phi_{G}\ge\phi$, and we call a partition $V=V_1 \sqcup\cdots\sqcup V_h$ 
a $\phi$-expander-decomposition if $\min_{i}\Phi_{G[V_{i}]}\ge\phi$.

\begin{theorem}[Theorem $1.2$ in~\cite{SW19}]
\label{thm:exp-dec}
Given a graph $G=(V,E)$ of $m$ edges and a parameter $\phi$, one can compute with high probability a partition $V=V_1 \sqcup\cdots\sqcup V_h$ such that $\min_i \Phi_{G[V_{i}]}\ge\phi$
and $\sum_{i}\delta(V_{i})=O(\phi m\log^{3}m)$.
In fact, the algorithm has a stronger guarantee that $\min_i \Phi_{G\{V_{i}\}}\ge\phi$.
The running time of the algorithm is $O(m\log^{4}m/\phi)$.
\end{theorem}

We will need the following strengthening of Theorem~\ref{thm:exp-dec},
where the input contains vertex demands (to be used instead of vertex degrees).
Given in addition a \emph{demand vector} $\dem\in \mathbb{R}_{\geq 0}^{V}$,
the graph $G=(V,E)$ is a \emph{$(\phi,\dem)$-expander} if for all subsets $S\subseteq V$, 
$\Phi_G^{\dem}(S) := \frac{\delta(S)}{\min(\dem(S),\dem(V\setminus S))} \geq \phi$. 
The following theorem statement is taken from \cite{LP20},
but its proof is actually a variation of a result from~\cite{CGLNPS20},
and gives a deterministic algorithm. 
We are not aware of a faster algorithm,
not even a randomized one based for instance on~\cite{SW19}.

\begin{theorem}[Theorem $III.8$ in \cite{LP20}]
\label{thm:exp-dec-dem}
Fix $\varepsilon>0$ and any parameter $\phi>0$.
Given an edge-weighted, undirected graph $G=(V,E,w)$
and a demand vector $\dem \in R_{\geq 0}^{V}$,
there is a deterministic algorithm running in time $O(m^{1+\varepsilon})$
that computes a partition $V=V_1 \sqcup\cdots\sqcup V_h$ such that
\begin{enumerate}
\item 
For each $i\in[h]$, define a demand vector $\dem_i\in \mathbb{R}^{V_i}_{\geq 0}$ given by $\dem_i(v)=\dem(v) + w(E({v}, V\setminus V_i))$ for all $v \in V_i$.
Then, the graph $G[V_i]$ is a $(\phi, \dem_i)$-expander.
\item 
The total weight of inter-cluster edges is
$w(E(V_1, \ldots, V_h)) = \sum_i w(E(V_i,V\setminus V_i)) \leq B\cdot \phi \dem(V)$
for $B = (\log n)^{O(1/\varepsilon^4)}$.
\end{enumerate}
\end{theorem}

\subsection{The Isolating Cuts Procedure}

A key tool that will be used in both our deterministic and randomized algorithms is a simple yet powerful procedure that can compute many minimum cuts all at once with $\tilde{O}(1)$ calls to a \MF algorithm.
More specifically, if a subset of the nodes $C \subseteq V$ is given, the procedure returns the minimum \emph{isolating} cut for each node $v \in C$; meaning the minimum cut that separates $v$ from all other nodes in $C \setminus \{v\}$.
This is very useful when we can pick a set $C$ that contains only one node from each minimum $p,v$-cut $C_v$ for many nodes $v$; then, all of these cuts are guaranteed to be found.

This lemma was introduced by Li and Panigrahy \cite{LP20} for their deterministic global minimum cut, and was later rediscovered in \cite{AKT21} (using a different proof and a slightly different statement where there is a pivot) and used in a \GHT algorithm.
Independently, Li and Panighrahy \cite{LP21} also used it for (an approximate) \GHT.

\begin{lemma}[The \pC Procedure (\cite{LP20}. See also~\cite{AKT21})]
\label{lem:proc_main}
Given an undirected graph $G=(V,E,c)$ on $n$ nodes and $m$ edges, a \emph{pivot} node $p\in V$, and a set of \emph{connected} vertices $C \subseteq V$, let $(C_v,V\setminus C_v)$ where $v \in C_v, p \in V\setminus C_v$ be the latest minimum $(p,v)$-cut for each $v \in C$.
One can compute deterministically in time $O(MF(n,m,c(E)) \cdot \log{n})$ $|C|$ disjoint sets $\set{ C'_v\subset V}_{v \in C}$ such that
$$
  \forall v\in C,
  \quad
  \text{if $C_{v} \cap C = \{v\}$ then $C'_v = C_v$.}
$$
\end{lemma}

\section{New Structural Results for the \GHT of Simple Graphs}
\label{sec:structural}


The main result of this section is a new structural result giving a non-trivial bound on a certain notion of depth of any \GHT of a simple graph $G$.
In fact, the bound is on the depth of a certain subtree of the \emph{cut-membership tree} defined next.

\begin{figure*}[!ht]
       \includegraphics[width=0.7\textwidth,center]{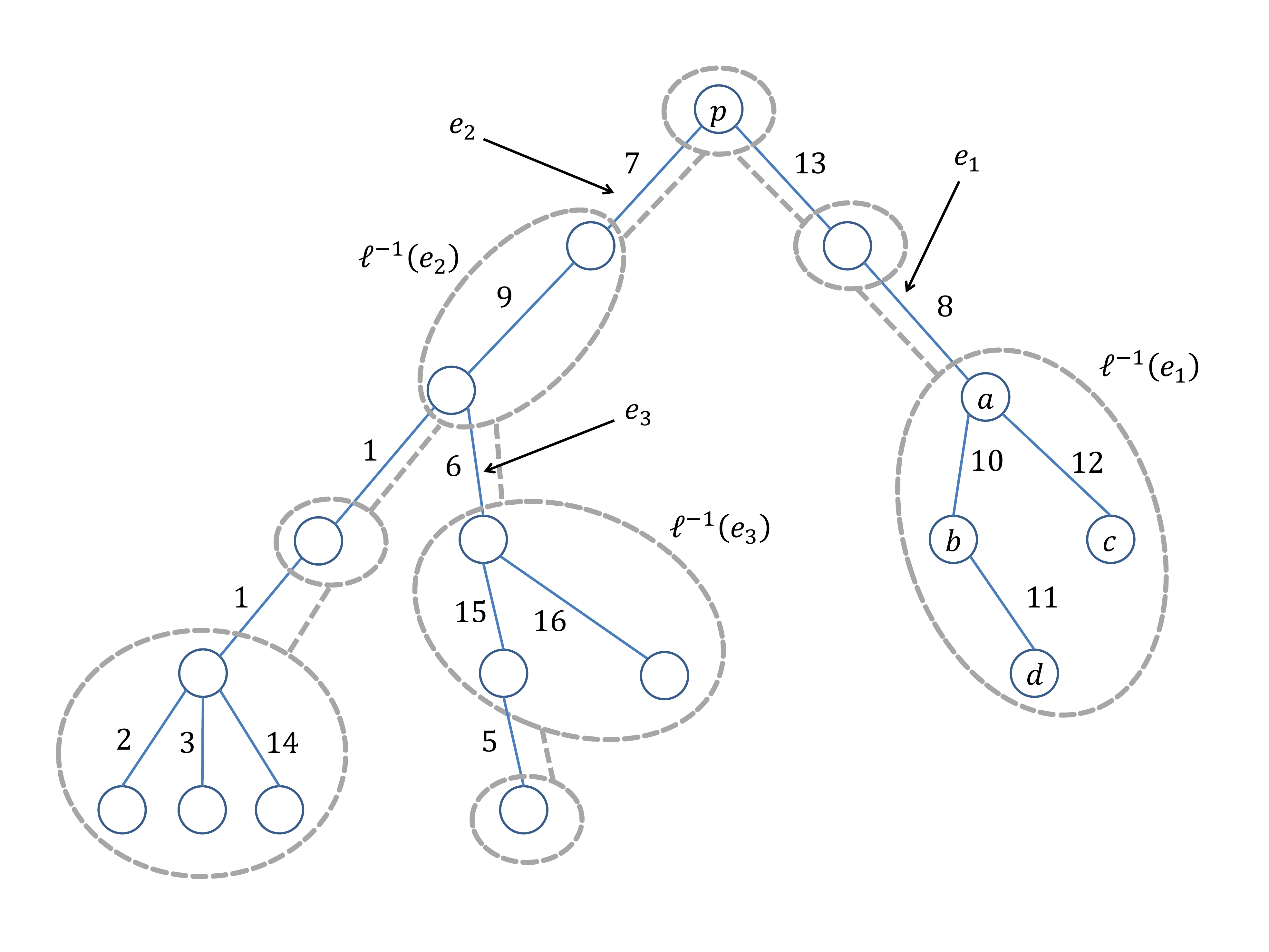}
   \caption[-]{
   An illustration showing a Gomory-Hu tree $\T$ with solid blue lines and the corresponding cut-membership tree $\TGP(G)$ with dashed gray lines. For example, $e_1=\ell(a)=\ell(b)=\ell(c)=\ell(d)$ and therefore $\{a,b,c,d\} \in V(\TGP)$ is a bag of size $4$. 
   }
   \label{Figs:Tzeta}
\end{figure*}

Let $\T$ be any \GHT of a graph $G$.
A cut-membership tree $\TGP$ with respect to node $p \in V(G)$ \cite[Section $3$]{AKT20_b} (see Figure~\ref{Figs:Tzeta} for illustration) is a coarsening of $\T$ where all nodes $B \subseteq V$ whose minimum cut to $p$ is the same are merged into one bag.
In more detail, define a function $\ell:V(\T)\setminus\set{p}\rightarrow E(\T)$, where $\ell(u)$ is the lightest edge in the path between $u$ and $p$ in $\T$, and $\ell(p)=\emptyset$;
to make the function well-defined we break ties by taking the lowest (furthest from $p$) minimal edge on the path.
Let $\TGP(G)$ (or simply $\TGP$ when $G$ is clear from the context) be the graph constructed from $\T$
by merging nodes whose image under $\ell$ is the same.
Observe that nodes that are merged together,
namely, $\ell^{-1}(e)$ for $e\in E(\T)$, are connected in $\T$,
and therefore the resulting $\TGP$ is a tree.
We shall refer to nodes of $\TGP$ as \emph{bags}.
For example, $p$ is not merged with any other node, and thus forms its own bag.
Just like a graph has many {\GHT}s, it also has many cut-membership trees with respect to any node $p$.

%

%

\begin{figure}[ht]
  \begin{center}
    \includegraphics[width=5.2in]{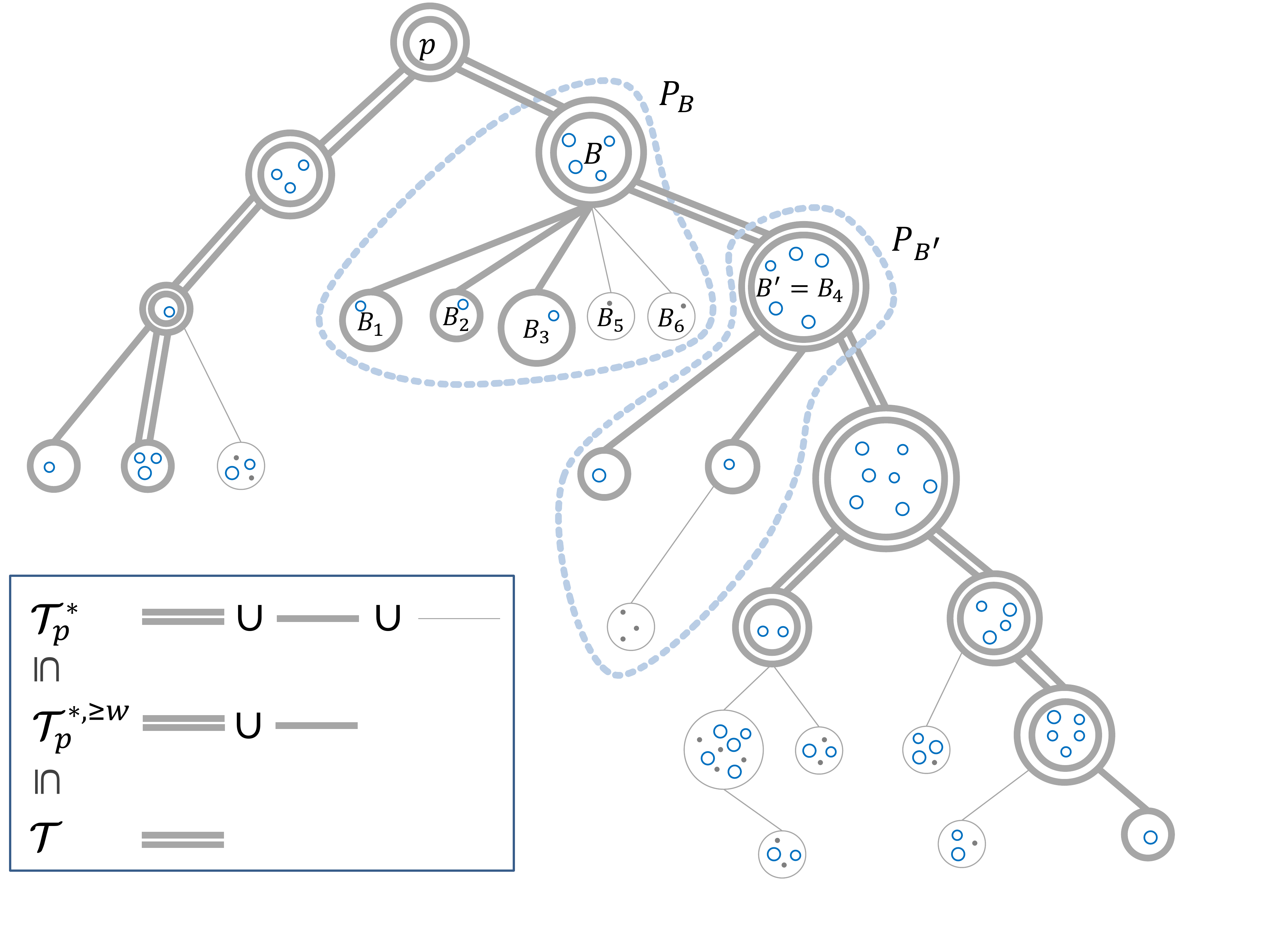}
    \end{center}
    \caption{An illustration of a cut-membership tree $\TGP$,  its subtree $\TGPW$ the $w$-large cut-membership tree, and its subtree $T$ that is obtained by deleting all easy bags. Small blue circles represent nodes of degree $\geq w$, while small grey dots represent nodes of degree $<w$}
    \label{Figs:Trees}
\end{figure}

The \emph{size} of a bag is defined to be the number of nodes (from $G$) that it contains.
Define the value of a bag to be the value of the minimum cut corresponding to it.
Define $\TGPW(G)$, called the $w$-large cut-membership tree of $G$,
to be the subset of $\TGP(G)$ containing only bags of value $\geq w$ (see Figure~\ref{Figs:Trees}).

Notice that all nodes in the bags of $\TGPW(G)$ have degree $\geq w$.
A bag of $\TGPW(G)$ is called \emph{easy} if the subtree under it contains only one node of degree $\geq w$.%
\footnote{Throughout, the subtree under a bag \emph{includes} the bag itself.}
It follows that all easy bags are leaves in $\TGPW(G)$, but not all leaves are easy.

\begin{theorem}
\label{thm:undone_bags}
%
If $\TGPW(G)$ is an arbitrary $w$-large cut-membership tree of a simple graph $G$ on $n$ nodes, then the number of bags in $\TGPW(G)$ that are not easy is $\leq c\cdot n/w$ for a fixed constant $c \leq 10^5$.
\end{theorem}

\begin{proof}
Let $\TGPW(G)$ be any $w$-large cut-membership tree of $G$, and define $T$ to be the tree $\TGPW(G)$ after the removal of its easy bags. 
Our goal is to bound the number of bags in $T$ by $O(n/w)$.
We begin by bounding its number of leaves, then continue to bound the number of internal bags (which is not straightforward since there could be many degree $2$ bags).

\begin{claim} \label{cl:Tleaves}
The number of leaves in $T$ is $\leq 2n/w$.
\end{claim}

\begin{proof}
Let $B_1,\ldots, B_x$ be the bags that are leaves in $T$.
Fix a node $v \in B_i$ and recall that the subtree of $\TGP(G)$ under $B_i$, when viewed as a subset of $V$, corresponds to a minimum $v,p$-cut in $G$.
Call the latter set $C_i \subseteq V$. 
It holds that the sets $C_1,\ldots,C_x$ are disjoint. 

The claim follows from the key observation that, since $G$ is simple, it must be that $|C_i| \geq w/2$ for all $i \in [x]$, which implies that $x \leq 2n/w$.
To see why this observation is true, assume for contradiction that $|C_i|<w/2$ for some $i \in [x]$.
Since $B_i$ is not an easy bag, the cut $C_i$ must contain at least two nodes of degree $\geq w$, call them $u,v$.
The node $u$ has $\geq w$ adjacent edges, but since $G$ is simple,
only $|C_i|-1<w/2-1$ of them could go into $C_i$,
and $\geq w/2+1$ of them must go to $V \setminus C_i$. 
This implies that $C_i \setminus \{u \}$ is a $p,v$-cut with fewer edges than $C_i$, a contradiction.
\end{proof}

Let $\calA\subset V(T)$ be the set of bags in $T$ whose degree (in $T$) is $2$.
By Claim~\ref{cl:Tleaves} and the basic combinatorial fact that in every tree,
the number of nodes of degree $>2$ is at most the number of leaves,
we obtain that $|V(T)\setminus \calA| \leq 2n/w$.
Thus, it remains to bound $|\calA|$.

Consider a bag $B \in \calA$. 
Let $B'$ be the bag below it in $T$, and let $P_B \subseteq V$ be the set of all nodes in the subtree under $B$ in $\TGP(G)$ except for those in the subtree under $B'$.
These nodes will be called the private nodes of $B$ since they do not appear in bags below $B$ in $T$.
For example, suppose that $B \in \calA$ has four children in $\TGPW(G)$ called $B_1,B_2,B_3,B_4$, out of which three are leaves $B_1,\ldots,B_3 \notin T$ and only one is internal $B'=B_4 \in T$, as well as two additional children $B_5,B_6$ that are in $\TGP(G)$ but not in $\TGPW(G)$; in this case, the set $P_B$ contains all nodes in $B_1,B_2,B_3,B_5,B_6$ and their subtrees in $\TGP(G)$, as well as the nodes in the bag $B$ itself.
Let ${\calA}_L \subseteq \calA$ be the set of bags of degree $2$ with a large private set ${\calA}_L = \{ B \in \calA \mid |P_B| > w/100 \}$.

Since for any two bags $B_1,B_2 \in \calA$ we have that $P_{B_1} \cap P_{B_2} = \emptyset$ it follows that $|{\calA}_L|\leq 100n/w$. 
The following key claim shows that $T'$ cannot have a long path of degree $2$ bags without large private sets.

\begin{claim}
\label{cl:10path}
There is no path of length $\geq 10$ in $T$ containing only bags from $\calA \setminus {\calA}_L$.
\end{claim}

\begin{proof}
Assume for contradiction that such a path exists,
and denote its bags by $B_1,\ldots,B_{10} \in \calA \setminus {\calA}_L$,
ordered by going away from $p$.
The nodes of $G$ can be partitioned into three sets $V=U \sqcup X \sqcup D$,
where $U$ contains all nodes that do not appear in bags in the subtree under $B_1$,
the set $X = P_{B_1} \cup \cdots \cup P_{B_{10}}$ of nodes that are private to bags in this path,
and $D$ contains all the remaining nodes, that is nodes that appear in the subtree under $B_{10}$ but not private to $B_{10}$.
Since none of these $10$ bags is in ${\calA}_L$ we have that $|X| \leq 10 \cdot w/100=w/10$.

For each $i \in [10]$,
let $\lambda_i$ be the value of the bag $B_i$ and fix a node $v_i \in B_i$.
By construction, $\lambda_1 \geq \cdots \geq \lambda_{10} \geq w$.
We know that $\deg(v_i) \geq \lambda_i \geq w$;
but since $|X|\leq w/10$ and $G$ is simple,
$v_i$ must have $>\lambda_i - w/10$ edges to nodes in $U \cup D$.
Overall, there must be $> \left(\sum_{i=1}^{10} \lambda_i \right)  - w> \sum_{i=1}^{9} \lambda_i$ edges from the nodes in $X$ to the nodes in $U \cup D$.
Now there are two cases:
either $|E(X,D)| \geq \lambda_{8}+\lambda_{9} \geq 2 \lambda_{10}$,
which contradicts the fact that $|E(V\setminus D,D)| \leq \lambda_{10}$;
or $|E(X,U)| > \lambda_{1}+\cdots + \lambda_{7}\geq \lambda_1 + w$,
which contradicts the fact that $|E(V\setminus U,U)| \leq \lambda_{1}$.
As both cases lead to a contradiction, Claim~\ref{cl:10path} follows. 
\end{proof}

Let $\calY$ be the set of bags in $T$ that are either leaves, or have degree $>2$, or have degree $2$ but are in ${\calA}_L$.
By the above and Claim~\ref{cl:Tleaves},
$|\calY| \leq 2\cdot 2n/w + |{\calA}_L| \leq 104n/w$.
To conclude the proof, we define a mapping $f: V(T)\setminus \calY \to \calY$
from bags that are yet to be counted to bags that have already been counted.
For a bag $B \in \calA \setminus {\calA}_L$, let $f(B)$ be the first bag in $\calY$ that is below $B$ in $T$.
By Claim~\ref{cl:10path}, there can only be $\leq 10$ bags in the pre-image of any bag in $\calY$.

To conclude the proof, we charge the remaining bags $V(T)\setminus \calY$
to the bags that we have already counted.
More precisely, we charge each bag $B \in V(T)\setminus \calY = \calA \setminus {\calA}_L$
to the first bag in $\calY$ that is below $B$ in $T$. 
By Claim~\ref{cl:10path}, at most $10$ bags are charged to the same bag in $\calY$,
and therefore $|\calA \setminus {\calA}_L| \leq 10 |\calY|$.
We conclude that $|V(T)| =  |\calA \setminus {\calA}_L| + |\calY| \leq 11 \cdot 104 \cdot n/w$,
which completes the proof of Theorem~\ref{thm:undone_bags}. 
\end{proof}

\section{Gomory-Hu Tree in Near-Quadratic Time}
\label{sec:NearQuadratic}

\begin{reptheorem}{thm1}
There is a randomized algorithm, with success probability $1-1/\poly(n)$, that 
constructs a \GHT of a simple graph $G$ and solves \APMF in time $n^{2+o(1)}$.
\end{reptheorem}

Recall from Section~\ref{prelim:GH} that an auxiliary graph $G'$ is obtained from a graph $G=(V,E)$ and from a \GHEPT $T$ by keeping the nodes $V' \subseteq V$ from one super-node of $T$ and contracting certain subsets of the other nodes as prescribed by $T$, introducing other nodes $V(G') \setminus V'$ into $G'$. 
The connectivity of any pair of original graph nodes $u,v \in V'$ in $G$ and $G'$ is the same (Lemma~\ref{lem:auxiliary}).
To simplify the exposition, we will use the following definition of connectivity of an auxiliary graph; note that the requirement is only for the original graph nodes $V'$ and not for the other nodes in $G'$.

\begin{definition}[Connectivity of an Auxiliary Graph]
We say that an auxiliary graph $G'$ has connectivity $k$
if for every pair of original graph nodes $u,v \in V'$, 
the minimum $(u,v)$-cut in $G'$ (and thus also in $G$) has value at least $k$.
\end{definition}

\subsection{Single-Source in Near-Quadratic Time}
\label{sec:SSGH}

\begin{theorem}
\label{thm:SSGH}
Given a simple graph $G=(V,E)$ on $N=|V|$ nodes with a designated pivot $p \in V$,
an auxiliary graph $G'$ on the graph nodes $V' \subseteq V(G)$ with $n=|V(G')|, m=|E(G')|$ and connectivity $\geq \sqrt{N}$,
and a perturbed version $\tilde{G}$ of $G'$ with unique minimum cuts,
one can compute the minimum $(p,v)$-cut in $\tilde{G}$ for all nodes $v \in V'$
in total time
$m^{1+o(1)} + Nn^{1+o(1)}$. 
\end{theorem}

In the rest of this section we prove Theorem~\ref{thm:SSGH}.
Denote the value of the minimum $u,v$-cut in $\tilde{G}$
by $\lambda_{u,v} := \MF_{\tilde{G}}(u,v)$, 
and fix a sufficiently large constant $\gamma\ge1$
(it will determine the overall success probability). 

\paragraph{The Algorithm}
First, initialize an estimate $c'(v) := \deg_{\tilde{G}}(v)$ for all $v \in V'$ along with a witness cut $C_v = \{v\}$.
Now for each $j = \lfloor \log{\sqrt{N}} \rfloor,\ldots, \lceil \log{N} \rceil$,
let $w:=2^j$ and execute the following process.
We shall refer to this process as stage $w$,
and it will compute correct estimates $c'(v)=\lambda_{p,v}$ (and witness cuts)
for all $v \in V'$ such that $w \leq \lambda_{p,v} < 2w$. 

\paragraph{Stage $\boldmath w$:}
\begin{enumerate} \compactify
\item
  Compute a Nagamochi-Ibaraki sparsifier $G_w$ (by Lemma~\ref{perturbed_NI})
  for the perturbed auxiliary graph $\tilde{G}$,
  so that this $G_w$ has $O(n w)$ edges,
  all cuts of value $< 2w$ are preserved,
  and all cuts of larger value still have value $\geq 2w$.

\item\label{easy_cuts}
  Construct the set $V'_{\ge w}:= \{ v \in V' \mid \deg_G(v) \geq w \}$
  and call the \pC procedure (see Lemma~\ref{lem:proc_main}) on $G_w,p,V'_{\ge w}$ to get a $p,v$-cut for all $v\in V'_{\ge w}$ and update the estimates $c'(v)$ and $C_v$ if the new cut has smaller value (and if its value is not $\geq 2w$).
  This step computes correctly all cuts that contain only one node from $V'_{\ge w}$; such cuts will be called easy below.%
  \footnote{Alas, if the true minimum cut is not easy,
    a returned cut containing one node from $S$ may be incorrect,
    and the algorithm will not know it.
  }

\item
  Initialize a set $C = \{ v \in V' \mid c'(v) > w \}$; its nodes will be called \emph{candidates}.

\item \label{it:SSGHrepeat} 
  While $|C| > \log{n}$ do:
  
\begin{itemize} \compactify
\item
  Compute an expander decomposition $(H_1,\ldots,H_\ell)$ of $G'$
  (see Theorem~\ref{thm:exp-dec-dem}) with
  parameters $\varepsilon:=(\log n)^{-1/9}=o(1)$
  and $\phi := 2^{- \log^{1/2}{n}} = n^{-o(1)}$, 
  and demand function 
\begin{equation}
    d(v) :=
    \begin{cases}
      w, & \text{if}\ v \in C \\
      0, & \text{otherwise.}
    \end{cases}
  \end{equation}

\item
  Define the \emph{size} of an expander $H$, denoted $\size_G(H)$,
  to be the total number of nodes from $G$ that appear in (the possibly contracted nodes) of $V(H) \subseteq V(G')$.%
  \footnote{For example, if $H$ contains two nodes from $V'\subseteq V(G)$ and two contracted nodes from $V(G') \setminus V'$ each of which obtained from ten nodes from $G$,
    then $\size_G(H) = 22$.
  }

\item
  For each expander $H_i$ with $\size_G(H_i) \geq w/2$ do:
\begin{enumerate}
\item \label{solve_x}
  Execute Procedure \pRighty (in Algorithm~\ref{alg:solve_x}).

\item \label{solve_y}
  Execute Procedure \pLefty (in Algorithm~\ref{alg:solve_y}).

\item Remove all nodes in $H_i$ from $C$.
\end{enumerate}

\end{itemize}

\item \label{all_C}
  Now that $|C| \leq \log{n}$, compute new cuts and estimates for all $v \in C$ using \MF invocations in $G_w$. 
\end{enumerate}


\begin{algorithm}
  \caption{Procedure \pRighty}\label{alg:solve_x}
  \SetKwFor{RepTimes}{repeat}{times}{end}
  \RepTimes{$2e\gamma \phi^{-1} \ln{N}$}{
    $S_{\geq w}\gets$ 
    randomly include each $v\in V(H_i) \cap C$ independently with probability $\phi$\\
    call the \pC procedure (Lemma~\ref{lem:proc_main}) on $G_w,p,S_{\geq w}$ 
    to get a $p,v$-cut $S'_v$ and a corresponding value $\delta(S'_v)$ for all $v\in S_{\geq w}$  \\
    \ForEach{ $v\in S_{\geq w}$ such that $\delta(S'_v) < \min\set{ c'(v), 2w }$ }{
      $c'(v)\gets \delta(S'_v) $\\
      set $S'_v$ as the witness cut for $v$
    }
  }
\end{algorithm}

\begin{algorithm}
  \caption{Procedure \pLefty}\label{alg:solve_y}
  \SetKwFor{RepTimes}{repeat}{times}{end}
  \SetKwFunction{ExtractMax}{ExtractMax}
  Initialize a heap containing the nodes of $H_i \cap C$ keyed by their estimate $c'(v)$ \\
  \RepTimes{$\alpha = 3\phi^{-1}$}{
    $v^*\gets$ node with maximum key $c'(v)$ extracted from the heap \\
    invoke \MF to obtain the minimum $p,v^*$-cut in $G_w$, call it $S^*$ and its value $\lambda_{p,v^*}$ \\
    \If{$c'(v^*) > \lambda_{p,v^*}$}{
      \texttt{$\alpha$++} 
      \tcp{increment the number of repetitions }
    }
    \ForEach{ $v\in S^*$ such that $c'(v)>\lambda_{p,v^*}$ }{
      $c'(v)\leftarrow \lambda_{p,v^*}$\\
      set $S^*$ as the witness cut for $v$
    }
    $C\leftarrow C\setminus \{v^*\}$\\
  }
\end{algorithm}

\paragraph{Running-Time Analysis}
We first bound the number of repetitions in Step~\ref{it:SSGHrepeat},
and then bound the time requried for each repetition.

\begin{claim}
Every repetition of Step~\ref{it:SSGHrepeat} reduces $|C|$ by at least factor $2$.
Consequently, this step is repeated at most $O(\log n)$ times (at a single stage $w$).
\end{claim}

\begin{proof}
Let $C$ be the set at the beginning of a repetition and let $C'$ be the set at its end.
The total demand $d(V)$ is exactly $|C| \cdot w$,
and therefore by Theorem~\ref{thm:exp-dec-dem} we have that
$B = (\log n)^{1/\varepsilon^4} = 2^{(\log\log n)(\log n)^{4/9}} \le O(\phi^{-1}/\log n)$
and the number of outer-edges is bounded by 
$$
  \sum_{i} |E(H_i,V \setminus H_i)|
  \leq O(d(V) \cdot B \cdot \phi)
  = O(|C|w/\log{n}).
$$
The set $C'$ contains all nodes in $C$ that appear in small expanders $H_i$ with $\size(H_i) < w/2$
(because nodes from the other expanders are removed from $C$).
Recall that the size of an expander is measured in $G$ even though the decomposition is of $G'$.
Any node $v \in C$ must have $\deg_G(v)+1 \geq \deg_{\tilde{G}}(v) \geq c'(v) \geq w$, and since $G$ is a simple graph, only $\size(H_i) \leq w/2$ of these edges can go to nodes in $H_i$.
Therefore, each node in $C'$ contributes $\geq (w-1)-w/2$ outer-edges to the expander-decomposition.
Thus, $|C'| \cdot (w/2-1) = O(|C|w/\log{n})$ and $|C'| < |C|/2$.

The $O(\log n)$ bounds on the number of repetitions follows immediately
because initially $|C| \leq |V'| \leq |V(G')| = n$. 
\end{proof}

\begin{claim}\label{claim:SS_time}
Each repetition in Step~\ref{it:SSGHrepeat} takes time 
$O(m^{1+\varepsilon}) = m^{1+o(1)}$
plus the time for
$O(\phi^{-1} N /w) = N /w \cdot n^{o(1)}$
\MF invocations in a graph with $n$ nodes and $O(nw)$ edges.
\end{claim}

\begin{proof}
By Theorem~\ref{thm:exp-dec-dem}, the expander-decomposition takes time
$O(m^{1+\varepsilon})$. 
The number of large expanders, i.e., with $\size(H_i) \geq w/2$,
is bounded by $O(N/w)$ since the number of nodes in $G$ is $N$
and each large expander contains a disjoint set of at least $w$ of them.

\begin{itemize}
\item
  In Step~\ref{solve_x}, for each of the $O(N/w)$ large expanders, there are $O(\log^{4} n\ln N)$ calls to the \pC procedure with the graph $G_w$ that has $O(nw)$ edges, each of which results in $O(\log{n})$ \MF invocations to a graph with $O(nw)$ edges.
\item
  The number of \MF invocations in Step~\ref{solve_y} is $\alpha = 3\phi^{-1}$ for each large expander,
  plus additional invocations across all the expanders. 
  Let $S_1,\ldots,S_k$ be the cuts that caused an increase in $\alpha$, 
  across all the expanders, listed in the order they were found.
  That is, each cut $S_i$ was obtained from a \MF invocation for some node $v_i$ (these were called $S^*$ and $v^*$ at the time),
  and before the invocation the estimate $c'(v_i)$ was strictly larger than the cut value that was found $\delta(S_i)=\lambda_{p,v_i}$.

  It follows that all these $k$ cuts are different.
  Indeed, Suppose that $S_i = S_j$ for $i<j$.
  Since the estimate of all nodes in $S_i$ were updated when $S_i$ was found,
  later, when $S_j$ was found, $c'(v_j)$ was already equal to $\delta(S_i)$
  and thus $S_j=S_i$ could not have improved $v_j$'s estimate, a contradiction.

  Moreover, and for a similar reason, all of these $k$ cuts are not easy.
  Recall that a minimum $p,v$-cut $S_v$ is called easy if $S_v$ contains only one node of degree $\geq w$.
  A cut that is easy would have been found by Step~\ref{easy_cuts},
  and thus could not have improved the estimate of any node.

  By Claim~\ref{Claim:rec_pert} below (proved using structural Theorem~\ref{thm:undone_bags}),
  the number $k$ of different cuts with these properties is bounded by $k\le O(N/w)$.
  Thus, the total number of \MF invocations in Step~\ref{solve_y} (Procedure Righty) is
  $O(N/w)\cdot 3\phi^{-1} + k = O(\phi^{-1} N /w)$,
  and each of them is made in the graph $G_w$ that has $O(nw)$ edges.
\end{itemize} 

Finally, the time it takes to check and update the estimates for all nodes $v \in S$ in a cut $S$ that is found, is always upper bounded by the time for the \MF invocation itself.
\end{proof}

Combining the above two claims we establish that the running time of stage $w$ of the algorithm is bounded by
$O(m^{1+\varepsilon}\log n) \leq m^{1+o(1)}$
plus the time for
$O(\phi^{-1} N /w\cdot \log n) \leq N /w \cdot n^{o(1)}$
invocations to a \MF algorithm on $n$ nodes and $O(nw)$ edges.
Since every stage has $w \geq \Omega(n^{1/2})$,
employing the recent $\tilde{O}(m+n^{1.5})$-time \MF algorithm \cite{linearflow21},
would solve each of these \MF invocations in time
$\tilde{O}(nw + n^{1.5}) \leq \tilde{O}(nw)$, 
which adds up over the entire stage to
$\tilde{O}(\phi^{-1} N /w\cdot \log n \cdot nw)
= \tilde{O}(Nn\cdot \phi^{-1}\log n )
\leq Nn^{1+o(1)}$. 
The running time of all steps other than~\ref{it:SSGHrepeat}
is dominated by $O(\log{N})$ \MF invocations in $G_w$,
which itself is negligible compared to the above.
We have $O(\log N)$ stages, and their overall time bound is 
$O(\log N)\cdot \tilde{O}(m^{1+\varepsilon}\log n + Nn\cdot \phi^{-1}\log n) 
 \leq \tilde{O}(m^{1+\varepsilon}+ Nn \phi^{-1})$.

To conclude the running time analysis,
it remains to prove Claim~\ref{Claim:rec_pert}
that was used to bound the number of additional invocations.
Recall that a minimum $(p,v)$-cut $S_v$ is called easy if $S_v$ contains $\leq 1$ nodes of degree $\geq w$.

\begin{claim}\label{Claim:rec_pert}
Let $G$ be a simple graph on $N$ nodes,
let $G'$ be an auxiliary graph of $G$ with respect to $V'\subset V(G)$ that includes a pivot $p$,
and suppose $\tilde{G}$ is a perturbed version of $G'$ with unique minimum cuts.
Then the number of different minimum $(p,v)$-cuts, over all $v \in V'$, that have value $\geq w$ and are not easy, is at most $O(N/w)$.
\end{claim}

\begin{proof}
Suppose that there is a set of $k>10^{5} \cdot N/w$ minimum $(p,v_i)$-cuts in $\tilde{G}$ (and therefore also in $G$) $C_1,\ldots,C_k$ for nodes $v_1,\ldots,v_k \in V'$.
Moreover, suppose that these cuts $C_1,\ldots,C_k$ are all distinct, have value $\geq w$ and are not easy.
Since $\tilde{G}$ has unique minimum cuts, we know that all these $k$ cuts are non-crossing.
Therefore, by Lemma~\ref{noncrossingGH}, there exists a \GHT $\T$ of $G$ that preserves all of these $k$ cuts.
Let $\TGP(G)$ be the cut-membership tree with respect to $p$ that corresponds to $\T$ and let $\TGPW(G)$ be the corresponding $w$-large cut membership tree.
It follows that all of our $k$ cuts must be preserved in $\TGPW(G)$ and, since they are different, they must correspond to different bags. Moreover, since the cuts are not easy, the bags they correspond to cannot be easy.
It follows that $\TGPW(G)$ contains $k>10^{5} \cdot N/w$ bags that are not easy, a contradiction to the structural Theorem~\ref{thm:undone_bags}.
\end{proof}

\paragraph{Correctness} 
We say that a node $v$ is done if $c'(v) = \lambda_{p,v}$, and our goal is to show that at the conclusion of the algorithm all nodes are done.
The argument is by induction on the stages. The follow main claim shows the inductive step.

\begin{claim}
\label{SS_correct}
Let $w=2^j$ for some $j \in \{\lfloor \log{\sqrt{N}} \rfloor,\ldots, \lceil \log{N} \rceil \}$ and consider the corresponding stage.
Suppose that all nodes $v \in V'$ with $\lambda_{p,v}<w$ are done at the beginning of the stage.
Then, with probability at least $1-1/N^{2\gamma}$, all nodes with $\lambda_{p,v}<2w$ are done at the end of the stage.
\end{claim}

\begin{proof}
Let $v \in V'$ be a node with $\lambda_{p,v}<2w$ that is not done at the beginning of the stage. 
It may be the case that $v$'s cut is easy and $v$ becomes done in Step~\ref{easy_cuts}.
If not, it must be that $v \in C$ when $C$ get initialized because $c'(v)>\lambda_{p,v}\geq w$.

There are two scenarios. Either $v$ is removed from $C$ in one of the repetitions of Step~\ref{it:SSGHrepeat} or it remains in $C$ until Step~\ref{all_C}.
The latter is an easy case because a direct \MF invocation is guaranteed to make $v$ done. (This is despite the fact that invocations are performed in $G_w$ rather than $G'$ because $G_w$ preserves all cuts of value $<2w$ in $\tilde{G}$.)
So suppose that $v$ was removed from $C$ due to being in a large expander $H_i$ during one of the repetitions of Step~\ref{it:SSGHrepeat}.

Let $(S_v, V \setminus S_v), v \in S_v, p \in V \setminus S_v$ be the minimum $p,v$-cut in $\tilde{G}$ (and therefore also in $G_w$).
Let $(L,R)$ be the projection of this cut unto the expander $H_i$, such that $L= H_i \cap S_v$ and $R=H_i \cap (V \setminus S_v)$.

By Theorem~\ref{thm:exp-dec-dem}, $H_i$ is a $(\phi,\dem_i)$-expander
for a demand vector $\dem_i \ge \dem$ (entrywise). 
Since $|E_{G'}(L,R)| \leq |E_{G}(S_v, V \setminus S_v)| <2w$
we can conclude that
$$
  \min( d(L), d(R))
  \leq \min( d_i(L), d_i(R))
  \leq \tfrac{ |E_{G'}(L,R)| }{\phi}
  \leq 2w \phi^{-1}. 
$$
By the definition of demand vector $\dem$,
we get that $\min( |L\cap C|, |R \cap C| ) \leq \phi^{-1}$,
where $C$ is the set of candidates at the beginning of the repetition of 
Step~\ref{it:SSGHrepeat} (and $v \in C$).
This means that we cannot have too many candidates in both sides,
and formally at least one of the following two cases (possibly both) must hold.

\begin{itemize}
\item Case $|L \cap C| \leq 2\phi^{-1}$.
  Then we claim that at least one of the sets $S$ that are chosen during the iterations of Step~\ref{solve_x} will be \emph{isolating} for $v$, meaning that $S \cap S_v = \{v \}$.
Once an isolating set $S$ is chosen, the call to the \pC procedure with $p,S,G_w$ is guaranteed to find the minimum $p,v$-cut and $v$ will be done.
To see this, let us compute the probability that $S$ is isolating for $v$ in one of the $2e\gamma \phi^{-1} \ln{N}$ iterations.
Since $v \in H_i \cap C$ the probability that it gets chosen to $S$ is exactly $\phi$.
The only other nodes from $S_v$ that can get chosen to $S$ are those in $L \cap C$, because we only choose nodes from $H_i \cap C=(L \cup R) \cap C$ and the nodes in $R$ are not in $S_v$. 
Thus, the probability that $v$ is chosen but not any of the other nodes in $S_v$ is at least $\phi\cdot (1-\phi)^{|L\cap C|}\geq \phi\cdot e^{-1}$.
Now, taking into account all the iterations, the probability that at least once an isolating set is chosen is $\geq 1- (1-\phi e^{-1})^{2e\gamma \phi^{-1}\ln{N}}\geq 1-1/N^{2\gamma}$.

\item Case $|R \cap C| \leq 2\phi^{-1}$.
  Then when we process $H_i$ in Step~\ref{solve_y} we are guaranteed to make $v$ done.
Suppose for contradiction that the step terminates while $v$ is still not done. 
First, notice that at least $3\phi^{-1}$ invocations for different nodes in $H_i \cap C$ are made during the step, while there are only $\leq \phi^{-1}$ nodes in $R \cap C$; therefore, there is at least one (in fact, at least $\phi^{-1}$) nodes $v^* \in L \cap C$ that are invoked.
Let $u$ be the last such node; this means that, when $u$ was chosen, it (1) must have had the largest estimate $c'(u)$ amongst all nodes in $H_i \cap C$, in particular, $c'(u)\geq c'(v)$; and (2) (since no more invocations were performed afterwards) it must have already been done before the invocation was performed, i.e. $c'(u)=\lambda_{p,u}$.
Thus, $c'(v) \leq c'(u) = \lambda_{p,u}$.
By our assumption that $v$ is not done, we have a third inequality: $c'(v) > \lambda_{p,v}$.
Combining it all together, we see that $\lambda_{p,v} < \lambda_{p,u}$ but this is a contradiction to the fact that $u \in L \subseteq S_v$: the cut $(S_v,V \setminus S_v)$ has value $\lambda_{p,v} < \lambda_{p,u}$ yet it separates $u$ from $p$, an impossibility by the definition of $\lambda_{p,u}$. 
\end{itemize}

This completes the proof the claim. 
\end{proof}

Finally, for the base case, it remains to argue that for the first stage, where $w=\sqrt{N}$, all nodes $v \in V'$ with $\lambda_{p,v}<w$ are done at the beginning of the stage.
This holds trivially because, by our assumption that the connectivity of the auxiliary graph $G'$ is $<\sqrt{N}$, there cannot be any node $v \in V'$ with $\lambda_{p,v}<w$.

\subsection{From Single-Source to All-Pairs via Randomized Pivot}
\label{sec:rand_pivot}

%
%

\paragraph{Overview}
The algorithm of the previous section computes the minimum cuts from a single pivot $p$ to all other nodes $v \in V$.
In this section, this algorithm is used recursively in order to construct a full tree in a manner that is similar to the Gomory-Hu algorithm (see Section~\ref{prelim:GH}).
Initially all nodes are in one super-node of an intermediate tree, and each time we refine the tree by partitioning the super-node into smaller super-nodes with the help of minimum cuts.
The Gomory-Hu algorithm partitions the super-node into two, while our algorithm (since it has all the minimum cuts from a single pivot) may divide it into many super-nodes at once.\footnote{This is a crucial point, as discussed in our previous papers \cite{AKT20,AKT20_b}}
For purposes of efficiency, we want to make the recursion depth logarithmic.
This is guaranteed if the cuts from the pivot happen to divide the graph in a balanced way; but unfortunately, it could be the case that the minimum cut from the pivot $p$ to all other nodes does not partition the graph well, e.g. it is $(\{p\},V\setminus\{p\})$.
A popular way to circumvent this issue (\cite{BHKP07,AKT20_b,AKT21}), that we also use in this section, is by picking the pivot at random and arguing that many of the cuts will be balanced, with high probability.
To facilitate such arguments, it is helpful to have unique minimum cuts in the graph; otherwise, a \MF invocation is free to always give us the most unbalanced cut with respect to the pivot we pick.
For this purpose, the algorithm of this section works with perturbed edge weights that make all minimum cuts unique.

\paragraph{The Algorithm}
The input is a simple graph $G$ on $N$ nodes.

\begin{enumerate}
\item Compute a partial $k$-tree $T_k$ for $G$ with $k=\sqrt{N}$ (see Section~\ref{prelim:kpartial}). This is the initial \GHEPT $T$ that will get refined throughout the algorithm until it is a \GHT of $G$. Note that from this point onwards, all auxiliary graphs will have connectivity $\geq k$.

\item For each super-node $V_i$ of $T_k$, do the following operations that refine $T$. 

 \begin{enumerate}
\item\label{recurse} If $|V_i|=1$, stop processing the super-node. Otherwise, continue:

\item Compute an auxiliary graph $G_i$ for $V_i$ using the current \GHEPT $T$ (see Section~\ref{prelim:GH}), and then compute a perturbed version $\tilde{G_i}$ for $G_i$ (as in Proposition~\ref{Proposition:Perturbation}) such that the minimum $p,v$-cuts for all $v\in V_i$ are unique. 

\item\label{random_pivot} Pick a pivot $p\in V_i$ uniformly at random, and run the algorithm of Theorem~\ref{thm:SSGH} on the graph $G$, the auxiliary graph $G_i$, its perturbation $\tilde{G_i}$, and the pivot $p$, to get the (unique) minimum $p,v$-cuts $S_v$ in $\tilde{G_i}$ for all $v \in V_i$,

\item Check how many nodes $v$ have ``good'' cuts. A cut $S_v$ is called good if the side of $v$ (not $p$) has $|S_v \cap V_i| \leq |V_i|/2$ graph nodes. The pivot $p$ is called good if all but $\leq 3|V_i|/4$ graph nodes in $V_i$ have good cuts.

\item\label{good_pivot} If the pivot $p$ is not good, go back to Step~\ref{random_pivot}. (If there is a sequence of $2\gamma \log{N}$ bad pivots in a row, abort and return failure.) Otherwise, continue:

\item Refine $T$ based on the cuts as follows. Assign each node $u \in V_i$ to the largest good cut $S_v$ that separates it from $p$, i.e. $u \in S_v$.\footnote{By largest we mean the cut with largest number of nodes $|S_v|$, not largest weight. This may not be a minimum cut for $u$ but it is a minimum $p,v$-cut for some $v \in V_i$ and therefore it can be used for refining $T$.} Let $S_1,\ldots,S_r$ be the (good) cuts that have at least one node $u \in V_i$ assigned to them.\footnote{Note that these cuts are non-crossing since they are minimum cuts in $\tilde{G}$ that has unique minimum cuts.} Refine $V_i$ by taking a Gomory-Hu step using each of these $r$ cuts.
This results in $r+1$ smaller super-nodes: one super-node $V_{i,j}$ for each cut $S_j$ plus a remainder super-node $V_{i,0}$ that contains $p$ and all other $\leq 3|V_i|/4$ nodes in $V_i$ whose cuts were not good (and therefore they were not assigned to any of the $r$ cuts).

\item Recurse on each of the smaller super-nodes $V_{i,j}$ for $j \in \{0,\ldots,r\}$ by going to Step~\ref{recurse}.

\end{enumerate}

\item When all super-nodes have been fully processed and contain only one node, the \GHEPT $T$ is a complete Gomory-Hu tree.
\end{enumerate}

\paragraph{Correctness}

Assuming that we do not reach a bad event in which we either abort in Step~\ref{good_pivot} or a perturbed graph does not have unique minimum cuts, the correctness of the algorithm is straightforward. 
It follows from the fact that whenever a super-node is partitioned into smaller super-nodes it is based on correct minimum cuts, and therefore the process could have been simulated by a Gomory-Hu algorithm that chose certain pairs and that received certain minimum cuts as invocation answers.
This is formalized in Lemma~\ref{lem:partial_tree_comb}, plus the observation that the partial $k$-tree $T_k$ is a \GHEPT (see Section~\ref{prelim:kpartial}).
The probability of the bad event can be upper bounded by $1/\poly(N)$ by Proposition~\ref{Proposition:Perturbation} and Claim~\ref{cl:clock} below, plus a union bound on the $O(N)$ recursive calls.

\begin{claim}
\label{cl:clock}
The probability that there is a sequence of $2\gamma \log{N}$ bad pivots is $\leq 1/N^{\gamma}$.
\end{claim}

\begin{proof}

We use Corollary $2.5$ from~\cite{AKT20_b}, given here for completeness.

\begin{corollary}\label{Corollary:Tournament}
Let $F=(V_F,E_F)$ be a graph where each pair of nodes $u,v\in V_F$ is associated with a cut $(S_{uv},S_{vu}=V_F\setminus S_{uv})$ where $u\in S_{uv}, v\in S_{vu}$ (possibly more than one pair of nodes are associated with each cut), and let $V'_F\subseteq V_F$. 
Then there exist $\card{V'_F}/2$ nodes $p'$ in $V'_F$ such that at least $\card{V'_F}/4$ of the other nodes $w \in V'_F{\setminus} \{p'\}$ satisfy $\card{S_{p'w}\cap V'_F}>\card{ S_{wp'} \cap V'_F }$. 
\end{corollary}

Assign $F:=G_i, V'_F=V_i$, and the cuts are the unique minimum cuts in the perturbed graph $\tilde{G_i}$.
It follows that there exist $\card{V_i}/2$ good pivots $p'\in V_i$, meaning that at least $\card{V_i}/4$ of the other nodes $q\in V_i\setminus \{p'\}$ satisfy that $S_q$, their minimum $p,q$-cut in $\tilde{G}$, is good.
Thus, the probability to not pick good pivots in all $2\gamma \log N$ iterations is at most $(1/2)^{2\gamma \log N}\leq 1/N^{2\gamma}$, as required.

\end{proof}

\paragraph{Running Time Analysis}
The algorithm is recursive. 
The depth of the recursion is $O(\log N)$ because each super-node on $n'$ graph nodes gets refined into smaller super-nodes containing at most $3n'/4$ graph nodes, starting from $n'=N$ and until $n'=1$.
The time for each recursive call is dominated by the logarithmic number of calls to Theorem~\ref{thm:SSGH} and is therefore $\tilde{O}(m+Nn)$ where $n$ and $m$ are the numbers of nodes and edges in the auxiliary graph that corresponds to the call.
To analyze the total time of the recursion, will use the fact that the total number of edges across all auxiliary graphs at a single depth is $O(M)$ (see Lemma $3.7$ in~\cite{AKT20_b}), and the total number of nodes there is $\tO(N)$ (see Lemma $4.1$ in~\cite{AKT21}). Thus, Summing over all the calls for one depth of the recursion gives 
$\tO(\sum_i m+Nn)\leq \tO(M+N^2)\leq \tO(N^2)$, and summing over all $\log N$ levels of the recursion concludes the total running time of $\tilde{O}(N^2)$.

\section{Derandomization via Dynamic Pivot}
 In this section we derandomize the algorithm of Section~\ref{sec:NearQuadratic}.
 The key tool is the dynamic pivot technique that will be presented in Section~\ref{sec:dynamic|_pivot} where we prove the following two theorems.
Our deterministic results for the single-source problem is the following; it is analogous to Theorem~\ref{thm:SSGH} in Section~\ref{sec:NearQuadratic}. 
 
 \begin{theorem}
\label{thm:SSGH_det}
Given a simple graph $G=(V,E)$ on $N$ nodes and an auxiliary graph $G'$ of $G$ on the graph nodes $V' \subseteq V$ with $n:=|V(G')|, m=|E(G')|$, one can find a certain pivot node $p \in V'$ and compute minimum $p,v$-cuts $S_v$ in $G$ for all terminals $v \in V'$ where $v\in S_v, p \in S_v$ and $|S_v \cap V'| \leq |V'|/2$, in $(m+Nn^{5/3})^{1+o(1)}$ deterministic time.
Suppose that single-pair \MF in $m$-edge weighted graphs can be solved deterministically in $m^{1+o(1)}$ time, then the time bound improves to $(m+Nn)^{1+o(1)}$.

\end{theorem}


Once we have this theorem, constructing the tree and proving Theorem~\ref{thm2} from the introduction is easy.
This is described in Section~\ref{sec:det_GH}.
Before going into the proof we begin with some preliminaries.

%
%
%

\subsection{Preliminaries for the Deterministic Algorithms}
\label{sec:prelimsDet}

\subsubsection{Latest Cuts}
\label{sec:latest}
Introduced by Gabow~\cite{Gabow91}, a \emph{latest} minimum $(s,t)$-cut with respect to $s$ (or a \emph{minimal} minimum $(s,t)$-cut, in some literature) is a minimum $(s,t)$-cut $(S_{s},S_{t}=V\setminus S_{s})$ such that no strict subset of $S_{t}$ is a minimum $(s,t)$-cut. The latest minimum cut (with respect to a node) is unique, and can be found in the same running time of any algorithm that outputs the maximum network flow between the pair, by finding all nodes that can reach $t$ in the residual graph.
In particular, the deterministic Goldberg-Rao algorithm for \MF also finds the latest minimum cut.

We will use the following basic lemma about latest cuts that follows from the submodularity of cuts.
Recall that a family of cuts is a laminar family if they are non-crossing: no pair $S,S'$ have both $S \setminus S' \neq \emptyset$ and $S' \setminus S \neq \emptyset$. 

\begin{lemma}
\label{lem:latest_noncrossing}
Let $p \in V$ be any node. 
The set of all minimum $p,v$-cuts with respect to $p$ for all $v \in V$ is a laminar family. 
\end{lemma}

 \subsubsection{Deterministic Hitting Sets via Splitters}
 \label{sec:splitters}

In order to derandomize the hitting-sets we pick for the \pC procedure, we employ \textit{splitters} (see~\cite{NaorSS95}). A splitter is a well-known combinatorial object previously applied in the areas pseudorandomness and derandomization (see, e.g. Section $5.6.1$ in~\cite{CyganFKLMPPS15}).

\begin{lemma}[See Lemma $III.4$ from~\cite{LP20}]
\label{lem:splitters}
Given a universe-size parameter $n$ and a size parameter $k$ one can deterministically construct a family of $s=\poly(k,\log{n})$ subsets $U_1,\ldots,U_s \subseteq [n]$ such that for any subset $T \subseteq [n]$ of size $|T|\leq k$ and a number $j\in T$ it holds that $U_i \cap T=\{j\}$ for some $i \in [s]$, in time $\tilde{O}(ns)$.
\end{lemma}

The proof of Lemma $III.4$ from~\cite{LP20} actually proves the stronger Lemma~\ref{lem:splitters}.

 \subsection{Single-Source Algorithm using a Dynamic Pivot}
 \label{sec:dynamic|_pivot}

\paragraph{The Algorithm}
Initially, set the pivot $p$ to be the node in $V'$ with largest degree (in $G$), and initialize an estimate $c'(v) := \deg_{\tilde{G}}(v)$ for all $v \in V'$ along with a witness cut $C_v = \{v\}$.
There are $O(\log{N})$ stages in the algorithm: the $w^{th}$ stage is guaranteed to compute correct estimates $c'(v)=\lambda_{p,v}$ for all $v \in V'$ such that $w \leq \lambda_{p,v} < 2w$, where $p$ is the pivot at the end of that stage.
Moreover, every estimate $c'(v)$ is witnessed by a \emph{good} $p,v$-cut $S_v,v\in S_v$ with $|S_v \cap V'| \leq |V'|/2$ and $\delta(S_v)=c'(v)$.

\begin{enumerate}

\item For each $w=2^j$ with $j \in \{1 \rfloor,\ldots, \lceil \log{N} \rceil \}$ do the following.
\begin{enumerate}
\item Compute a Nagamochi-Ibaraki sparsifier $G_w$ of $G'$ with $O(n w)$ edges and such that all cuts of value $< 2w$ are preserved, and all larger cuts still have value $\geq 2w$. 

\item\label{easy_cuts_det} Construct the set $V'_{\geq w}:= \{ v \in V' \mid \deg_G(v) \geq w \}$ of degree $\geq w$ nodes, and call the \pC procedure (Lemma~\ref{lem:proc_main}) on $G_w,p,V'_{\geq w}$ to get a $(p,v)$-cut $S_v$ for all $v\in V'_{\geq w}$ such that the sets $\{S_v\}_{v \in S}$ are disjoint.
If there is a node $q \in V'_{\geq w}$ such that $|S_q \cap V'| > |V'|/2$, invoke a \MF algorithm to find the latest minimum $p,q$-cut with respect to $p$,\footnote{That is, the minimum $p,q$-cut in which the side of $q$ is as small as possible.} and denote it by $S_q$.\footnote{The previous $S_q$ can be discarded.}
If it is still the case that $|S_q \cap V'| > |V'|/2$, perform the \emph{pivot-change protocol} (described below) making $q$ the pivot instead of $p$.
Finally, for each $v \in V'_{\geq w}$ update the estimate $c'(v)$ and the witness $C_v$ if the new cut $S_v$ has smaller value.

\item Initialize the set $C = \{ v \in V' \mid c'(v) > w \}$.
\item \label{repeat_det}   While $|C| > \log{n}$ do:
 
\begin{itemize}
\item   Compute an expander decomposition $(H_1,\ldots,H_\ell)$ of $G'$
  (see Theorem~\ref{thm:exp-dec-dem}) with 
  parameters $\varepsilon:=(\log n)^{-1/9}$
  and $\phi := 2^{- \log^{1/2}{n}}$, 
and demand function:
\begin{equation}
    d(v)=
    \begin{cases}
      w, & \text{if}\ v \in C \\
      0, & \text{otherwise.}
    \end{cases}
  \end{equation}

\item   Define the \emph{size} of an expander $H$, denoted $\size_G(H)$,
  to be the total number of nodes from $G$ that appear in (the possibly contracted nodes) of $V(H) \subseteq V(G')$.%
\item For each expander $H_i$ such that $\size(H_i) \geq w/2$ do:
\begin{enumerate}
\item\label{solve_x_det} (This corresponds to Procedure \pRighty in Section~\ref{sec:SSGH}.) Use \emph{splitters} (Lemma~\ref{lem:splitters}) with universe-size parameter $u=|H_i \cap C|$ and size parameter $\ell=2\phi^{-1}$ to obtain a family of sets $U_1,\ldots,U_s \subseteq [u]$.\footnote{In the notation of Lemma~\ref{lem:splitters}, $u$ is $n$ and $\ell$ is $k$.} For each $j \in [s]$ use $U_j$ to do the following: Construct a set $S_{\geq w} \subseteq V'$ by adding the $x^{th}$ node in (a lexicographic ordering of) $H_i \cap C$ to $S_{\geq w}$ if and only if $x \in U_j$, and call the \pC procedure (Lemma~\ref{lem:proc_main}) on $G_w,p,S_{\geq w}$ to get a $(p,v)$-cut $S_v$ for all $v\in S_{\geq w}$ such that the sets $\{S_v\}_{v \in S_{\geq w}}$ are disjoint.
If there is a node $q \in S_{\geq w}$ such that $|S_q \cap V'| > |V'|/2$, invoke a \MF algorithm to find the latest minimum $p,q$-cut with respect to $p$ and denote it by $S_q$.
If it is still the case that $|S_q \cap V'| > |V'|/2$, perform the \emph{pivot-change protocol} making $q$ the pivot instead of $p$.
Finally, for each $v \in S_{\geq w}$ update the estimate $c'(v)$ and the witness $C_v$ if the new cut $S_v$ has smaller value.

\item \label{solve_y_det} (This corresponds to Procedure \pLefty in Section~\ref{sec:SSGH}.) Store the nodes in $H_i \cap C$ in a data structure (e.g. a heap) that allows to extract the node with largest $c'(v)$, and to decrease $c'(v)$ for any node. Perform the following operation $\alpha = 3\phi^{-1}$ times:
\begin{enumerate}
\item \label{repeated_query_det} Extract $v^*$ the node in $H_i \cap C$ with largest $c'(v)$. Invoke a \MF algorithm to obtain the latest minimum $p,v^*$-cut in $G_w$ with respect to $p$, call it $S^*, v^* \in S^*$ and its value $\lambda_{p,v^*}$.
If $c'(v^*) > \lambda_{p,v^*}$, the number of repetitions to Step~\ref{repeated_query_det} is increased by one, namely \texttt{$\alpha$++}.
If $|S^* \cap V'|> |V'|/2$, perform the \emph{pivot-change protocol} making $v^*$ the pivot instead of $p$.
    For each $v\in S^*$ such that $c'(v)>\lambda_{p,v^*}$ do $c'(v)\leftarrow \lambda_{p,v^*}$ and set $S^*$ as the witness cut for $v$.
Remove $v^*$ from $C$.
\end{enumerate}

\item Remove all nodes in $H_i$ from $C$.
\end{enumerate}

 \end{itemize}
 \item \label{all_C_det} Once $|C| \leq \log{n}$, compute the latest minimum $p,v$-cut $S_v$ for all $v \in C$ by directly invoking a \MF algorithm in $G_w$, and update the estimates accordingly. 
 If any of the invocations for a node $q \in C$ results in a cut $S_q$ with $|S_q \cap V'| > |V'|/2$ then the algorithm calls the \emph{pivot-change protocol} making $q$ the pivot instead of $p$.

\end{enumerate}

\end{enumerate}

\paragraph{The pivot-change protocol}
A pivot change from $p$ to $q$ takes place when the algorithm finds out that the latest minimum $p,q$-cut $S_{p,q}, p \in S_{p,q}$ with respect to $p$ is not good, i.e. $|S_{p,q} \cap V' |< |V'|/2$.
Since it is latest, it implies that the side of $p$ in \emph{all} minimum $p,q$-cuts will contain less than $1/2$ of the nodes of $V'$.
Recall that $\lambda_{p,q} = \delta(S_{p,q})$.
To change the pivot from $p$ to $q$ the algorithm performs the following operations.

\begin{itemize}

\item Let $S_{q,p},p \in S_{q,p}$ be the latest minimum $p,q$-cut with respect to $q$.\footnote{In other words, among the minimum $p,q$-cuts, $S_{p,q}$ was the cut maximizing the side of $p$ and $S_{p,q}$ is the cut that minimizes it. Note that $|S_{q,p} \cap V' | \leq |S_{p,q} \cap V' |< |V'|/2$.}

\item Go over all nodes $v \in V'$ and if $c'(v) > \lambda_{p,q}$ set $c'(v) := \lambda_{p,q}$ with the witness cut being $S_{q,p}$.
 
\end{itemize}

The running time for the pivot change protocol is the time of a \MF algorithm plus $O(n)$ because we spend $O(1)$ time for each node $v \in V'$.

The key claim is the following, showing that the algorithm can be resumed after a pivot-change protocol and it will be as if $q$ was the pivot all along: any node that already had a correct good cut with respect to the old pivot $p$ will also have a correct good cut with respect to the new pivot $q$.

\begin{claim}
\label{pchange_correct}
Suppose that, before the pivot-change protocol, the estimate of node $v \in V'$ was correct, i.e. $c'(v)=\lambda_{p,v}$, and its witness cut  was good, i.e. $S_v, v\in S_v$ and $|S_v \cap V'| \leq |V'|/2$, then after the pivot-change protocol it holds that $c'(v)=\lambda_{q,v}$ and its witness cut $S_v', v \in S_v'$ has $|S'_v \cap V'| \leq |V'|/2$.
Moreover, if the witness cut before the change was latest with respect to $p$ then the cut after the change is also latest with respect to $q$.
\end{claim}

\begin{proof}
Let $v$ be such a node and let $S_1, v \in S_1$ be its witness cut before the protocol. 
That is, $S_1$ is a minimum $p,v$-cut with $\delta(S_1)=\lambda_{p,v}$ and $|S_1 \cap V'| \leq |V'|/2$.
Let $S_2$ be $S_{q,p}, p \in S_{q,p}$ the cut that is used as an alternative in the pivot-change protocol.
That is, $S_2$ is the latest minimum $p,q$-cut with respect to $q$ with $\delta(S_2)=\lambda_{p,q}$.

Our goal is to show that one of $S_1$ or $S_2$ is a minimum $q,v$-cut and that it is the one with smaller value. (And then also to show that if $S_1$ was latest with respect to $p$ then the one with smaller value is also latest with respect to $q$.)
There are two cases:

\begin{itemize}
\item If $v \in S_2$ then both $S_1,S_2$ are $q,v$-cuts and both contain $\leq 1/2$ of the nodes of $V'$; the algorithm will assign the one with smaller value and our goal is to show that it is optimal.
Suppose for contradiction that the minimum $q,v$-cut $S', v \in S'$ has smaller value than $\min(\delta(S_1),\delta(S_2))$.
The node $p$ is either in $S'$ or outside it, both cases are impossible.
If $p \in S'$ then $S'$ is a $p,q$-cut with value smaller than $\delta(S_2)$, and if $p \notin S'$ then $S'$ is a $p,v$-cut with value smaller than $\delta(S_1)$.
In the same way, suppose for contradiction that the latest minimum $q,v$-cut $S''$ with respect to $q$ has fewer nodes than the cut we have chosen.
If $p \in S''$ then it is a violation to the fact that $S_2$ is latest, and if $p \notin S''$ then it is also a violation because $S''$ is a minimum $p,v$-cut with fewer nodes than $S_1$.

\item If $v \notin S_2$ then it must be that $\delta(S_1) \leq \delta(S_2)$, in which case the algorithm will assign $S_1$.
First, observe that $S_1$ is indeed a $q,v$-cut; otherwise, if $q \in S_1$, it would be a violation to the condition of a pivot-change because $S_1$ would be a minimum $p,q$-cut with $|S_1 \cap V'| \leq |V'|/2$.
%
%
Second, to show the optimality of $S_1$, suppose for contradiction that the minimum $q,v$-cut $S', v \in S'$ has $\delta(S')<\delta(S_1)\leq \delta(S_2)$.
It follows that $p \in S'$ or else $S'$ is a $p,v$-cut with value smaller than $\delta(S_1)$.
But then $S'$ is a $p,q$-cut with value smaller than $\delta(S_2)$, a contradiction.
Similarly, suppose for contradiction that the latest minimum $q,v$-cut $S''$ has fewer nodes than $S_1$.
Again, $p \in S''$ or else it violates the fact that $S_1$ is the latest minimum $p,v$-cut with respect to $p$, but then $S''$ contains a node that $S_1$ does not contain (the node $p$) despite having the same value, a contradiction to it being latest.

\end{itemize}

\end{proof}

\paragraph{Running Time Analysis}
The running time analysis is identical to that of Section~\ref{sec:SSGH} with the following minor changes. 
\begin{itemize}
\item The deterministic algorithm may invoke the pivot-change protocol after making a call to the \pC procedure or after invoking a \MF algorithm. This does not increase the total running time because the running time of the protocol is on the order of the time for a single \MF invocation.

\item In the proof of Claim~\ref{claim:SS_time}, the number of additional invocations made in Step~\ref{solve_y} is upper bounded by $\tilde{O}(N/w)$ by showing that each such invocation on node $v_i$ can be associated with a cut $S_i$ and then bounding the number of such cuts using Claim~\ref{Claim:rec_pert}. 
The latter claim used the fact that these cuts were computed in a perturbed graph in order to conclude that they are non-crossing.
In the deterministic algorithm, this use of the perturbation is replaced by working with latest cuts (this is only important in Step~\ref{solve_y_det}).
Indeed, all of the cuts $S_i$ that are associated with additional invocations are latest cuts with respect to a pivot $p$ and this allows us to conclude that they are non-crossing (by Lemma~\ref{lem:latest_noncrossing}) and the same proof goes through.
However, a potential issue is that these cuts might be latest with respect to different pivots due to the pivot changes.
To resolve this issue, Claim~\ref{pchange_correct} proves that after the pivot-change (from $p$ to $q$) any cut that was latest with respect to $p$ is also latest with respect to $q$.

\item The use of splitters (Lemma~\ref{lem:splitters}) instead of a randomized hitting set only incurs additional $n^{o(1)}$ factors.

\item Finally, the most consequential change is due to the difference in complexity between the current deterministic and randomized \MF algorithms.
In stage $w$ the number of edges in $G_w$ is $m=O(nw)$ and the time for the $\tilde{O}(N/w)$ invocations is $\tilde{O}(N/w \cdot nw \cdot n^{2/3}) = \tilde{O}(N n^{5/3})$ using the $\tilde{O}(|E| \cdot |V|^{2/3})$ bound of Goldberg and Rao~\cite{GR98} and it is $N/w \cdot (nw)^{1+o(1)}=N n^{1+o(1)}$ assuming an almost-linear time \MF algorithm.


\end{itemize}

\paragraph{Correctness} The same correctness proof as in Section~\ref{sec:SSGH} works here as well, due to the fact that the pivot-change protocol makes it as if the new pivot was the pivot from the start (as is formally proven in Claim~\ref{pchange_correct}).
The only minor difference is in the proof of Claim~\ref{SS_correct} where instead of arguing that the randomized process in Step~\ref{solve_x}  succeeds  in finding an isolating set (for any given node $v$) with high probability, we use Lemma~\ref{lem:splitters} that guarantees the success of Step~\ref{solve_x_det} with a deterministic construction.

\subsection{From single-source to all-pairs}
\label{sec:det_GH}

After we have the algorithm of Theorem~\ref{thm:SSGH_det}, constructing a \GHT becomes easy.

\begin{reptheorem}{thm2}[Restated]
There is a deterministic algorithm that constructs a \GHT of a simple graph
and solves \APMF in time $\tilde{O}(n^{2\frac{2}{3}})$. 
The time bound improves to $n^{2+o(1)}$ assuming that single-pair \MF can be computed in deterministic time $m^{1+o(1)}$ in $m$-edge weighted graphs.
\end{reptheorem}

\paragraph{The Algorithm}
The input is a simple graph $G$ on $N$ nodes.

\begin{enumerate}
\item Start from an intermediate tree $T$ containing one super-node $V$, and do the following (recursive) operations to refine it.

 \begin{enumerate}
\item\label{recurse_det} If $|V_i|=1$, stop processing the super-node. Otherwise, continue:

\item Compute an auxiliary graph $G_i$ for $V_i$ using the current \GHEPT $T$ (see Section~\ref{prelim:GH}).

\item\label{random_pivot_det} Run the algorithm of Theorem~\ref{thm:SSGH_det} on the graph $G$ and the auxiliary graph $G_i$, to get a pivot $p \in V_i$ and minimum $p,v$-cuts $S_v$ for all $v \in V_i$, all of which are good in the sense that $|S_v \cap V_i| \leq |V_i|/2$.

\item Refine $T$ based on the cuts as follows. Assign each node $u \in V_i$ to the largest cut $S_v$ that separates it from $p$, i.e. $u \in S_v$,  breaking ties in an arbitrary but consistent manner. Let $S_1,\ldots,S_r$ be the cuts that have at least one node $u \in V_i$ assigned to them. Observe that all nodes in $V_i \setminus \{p\}$ are contained in exactly one of these cuts. Refine $V_i$ by taking a Gomory-Hu step using each of these $r$ cuts.
This results in $r+1$ smaller super-nodes: one super-node $V_{i,j}$ for each cut $S_j$ plus a remainder super-node $V_{i,0}$ only containing $p$.

\item Recurse on each of the smaller super-nodes $V_{i,j}$ for $j \in \{0,\ldots,r\}$ by going to Step~\ref{recurse_det}.

\end{enumerate}

\item When all super-nodes have been fully processed and contain only one node, the \GHEPT $T$ is a complete Gomory-Hu tree.
\end{enumerate}

\paragraph{Correctness} The correctness follows directly from the correctness of the Gomory-Hu framework (Section~\ref{prelim:GH}) and the correctness of the single-source algorithm that we use (Theorem~\ref{thm:SSGH_det}).

\paragraph{Running Time} The analysis of the recursion is very similar to that of Section~\ref{sec:rand_pivot}. The depth is logarithmic due to ``goodness'' of the cuts returned by the dynamic pivot technique. The total number of nodes in all auxiliary graphs of each level is $O(N)$, which makes the total running time of the level $N^{2+o(1)}$ if a near-linear time \MF algorithm is assumed or $N^{8/3+o(1)}$ unconditionally (using Theorem~\ref{thm:SSGH_det}).

\section{A Barrier of $\Omega(n^{1.5})$}


\begin{reptheorem}{thm:reduction}
Assuming Hypothesis~\ref{nonsimple}, there is an $n^{1.5-o(1)}$ lower bound for computing a Gomory-Hu tree of a simple graph on $n$ nodes and $O(n)$ edges. 
\end{reptheorem}

\begin{proof}
Assume for contradiction that there is an $n^{1.5-\varepsilon+o(1)}$ time algorithm for computing a Gomory-Hu tree of a simple graph on $n$ nodes and $O(n)$ edges. We will show that it implies an $n^{3-\varepsilon'+o(1)}$ time algorithm for unweighted multigraphs, refuting Hypothesis~\ref{hypo1}. 
Let $G=(V,E)$ be an unweighted multigraph with $\card{E}=O(\card{V}^2)$. Subdivide each edge $(u,v)\in E$ by placing exactly one new node $l_{uv}$ along the edge, and denote the new graph $G'=(V',E')$. Observe that the resulting graph is simple, and has $\card{V'}\leq O(\card{V}^2)$ nodes, and $\card{E'}\leq O(\card{V'})$ edges.
Using the assumed algorithm on $G'$, we get a cut-equivalent tree $T'$ for $G'$ in a total running time of $\card{V'}^{1.5-\varepsilon+o(1)}\leq n^{3-\varepsilon'+o(1)}$ for some $\varepsilon'=2\varepsilon>0$.

Next, using $T'$ we construct a cut equivalent tree $T$ for $G$, as follows. 
Recall that Gusfield's algorithm~\cite{Gusfield90} can construct a cut-equivalent tree by calling $n-1$ times to a \MF routine on the input graph.
Apply Gusfield algorithm on $G$, such that whenever a pair $s,t$ is invoked, find the the partition of $V'$ according to $T'$, and output its projection on $V$ (that is, output the partition ignoring the added nodes in $V'\setminus V$).

Observe that a minimum-cut in $G'$ between every two nodes $u,v\in V'$ that have copies in $V$ projected onto $V$ is a minimum $u,v$-cut in $G$, and vice-versa.
First, a minimum $u,v$-cut in $G'$ has either $(u,l_{uv})$ or $(v,l_{uv})$ (but not both) in the cut: if both are in the cut, then by moving $l_{uv}$ to the other side we can decrease the cost of the cut by $2$, a contradiction. As a result, the cost of the minimum $u,v$-cut in $G'$ is precisely the same as the cost of the projection in $G$.
Second, trivially, a minimum cut in $G$ has a cut in $G'$ that partitions original nodes the same way and is of the same cost, as required.

\end{proof}
 
\section{Conclusion}

The main result of this paper is an $n^{2+o(1)}$ time algorithm for constructing the \GHT of an unweighted graph, leading to an almost-optimal algorithm for \APMF in unweighted graphs.
Even assuming an almost-linear time \MF algorithm (as in Hypothesis~\ref{hypo1}) the previously known algorithms \cite{AKT21} could not go below $n^{2.5}$.
The improvement is achieved by utilizing a stronger primitive than previous work: an expander decomposition with \emph{vertex demands}. 
Fitting this tool into the Gomory-Hu framework involves several technical challenges, and a new structural understanding of the \GHT of a simple graph.
To reach the $n^{2+o(1)}$ bound, the algorithm exploits the simplicity of the graph in a second way compared to the $n^{2.5}$ algorithm, and therefore it may be even harder to adapt for weighted graphs.
The main open question in the field has become even more outstanding.

\begin{oq}
\label{mainoq}
Can the \GHT of a weighted graph be constructed in subcubic time?
\end{oq}

The first step towards this question would be to achieve a subcubic algorithm for unweighted \emph{multigraphs} on $O(n^2)$ edges; i.e. to refute Hypothesis~\ref{nonsimple}.
Interestingly, due to Theorem~\ref{thm:reduction}, this is required before an $m^{1+o(1)}$ algorithm can be achieved in simple graphs.

We also introduce a new derandomization technique that replaces the randomized pivot selection of recent works with a dynamic pivot.
It gives hope that whatever bounds are achieved with randomized algorithms, in the context of \GHT algorithms, can also be derandomized.

Finally, an interesting take-away message from this paper is that \APMF suddenly appears to be an easier problem than \APSP.
At least in unweighted graphs, and at least with current algorithms, its upper bound is $n^{2+o(1)}$ compared with the $n^{\omega+o(1)}$ for \APSP.
The relative easiness is not only theoretical: the methods used by our \APMF algorithm are much less infamous than the fast matrix multiplication that goes into the \APSP algorithm.
However, formally proving a separation between the two is difficult, as it requires proving an $\Omega(n^{2+\eps})$ lower bound for matrix multiplication, and may not even be possible because $\omega$ is often conjectured to be $2$.
Perhaps the more promising direction is by resolving Open Question~\ref{mainoq} with an $O(n^{3-\eps})$ time algorithm for \APMF in weighted graphs.
This would establish a separation, assuming the popular conjecture that \APSP in weighted graphs requires $n^{3-o(1)}$ time \cite{VW10}, making Open Question~\ref{mainoq} even more consequential.

\medskip
\paragraph{Acknowledgements}
We are grateful to Thatchaphol Saranurak for helpful clarifications regarding the literature on randomized versus deterministic expander decompositions.

{\small
\bibliographystyle{alphaurlinit}

\bibliography{robi}
}

\appendix

\newpage

\end{document}